\documentclass[onecolumn,aps,prd,preprintnumbers,showpacs,superscriptaddress,nofootinbib,amsmath,amssymb,floats,floatfix,showkeys,notitlepage,longbibliography]{revtex4-1}

\usepackage{orcidlink}
\usepackage{comment}
\usepackage[utf8]{inputenc}
\usepackage[T1]{fontenc}
\usepackage{lipsum}
\usepackage{graphicx}
\usepackage{subfigure}
\usepackage{palatino}
\usepackage{sans}
\usepackage{hyperref}
\hypersetup{colorlinks=true,linkcolor=blue,urlcolor=blue,citecolor=blue}
\usepackage[toc,page]{appendix}
\usepackage[normalem]{ulem}
\usepackage{adjustbox}
\usepackage{latexsym}
\usepackage{amsmath}
\usepackage{amssymb}
\usepackage{amsfonts}
\usepackage{dcolumn}
\usepackage{bm}
\usepackage{tikz}
\usepackage{bigints}
\usepackage{array,tabularx,multirow,booktabs}
\usepackage[tracking=true]{microtype}
\usepackage{soul} %for highlighting
\usepackage{booktabs}
\usepackage{multirow}
\SetTracking{}{500}
\SetTracking{encoding={*}, shape=sc}{40}
\UseRawInputEncoding %for inputenc error%
\allowdisplaybreaks

\begin{document} \sloppy
%\title{Light deflection angle in generic static spherically symmetric spacetimes using homotopy perturbation and variational iteration methods}

\title{Semi-Analytic Trajectory Analysis of Light in Generic Static Spacetimes}

\author{Ali \"Ovg\"un \orcidlink{0000-0002-9889-342X}}
\email{ali.ovgun@emu.edu.tr}
\affiliation{Physics Department, Eastern Mediterranean University, Famagusta, 99628 North
Cyprus via Mersin 10, Turkiye.}

\author{Reggie C. Pantig \orcidlink{0000-0002-3101-8591}} 
\email{rcpantig@mapua.edu.ph}
\affiliation{Physics Department, School of Foundational Studies and Education, Map\'ua University, 658 Muralla St., Intramuros, Manila 1002, Philippines.}

\begin{abstract}
We study a unified semi-analytical framework to study null geodesics and
weak-field light deflection in generic static, spherically symmetric
spacetimes of the form
\(ds^2 = -\alpha(r,\delta)\,dt^2 + \gamma(r,\delta)\,dr^2 +
\beta(r,\delta)\,d\Omega_2^2,\)
where $\alpha$, $\beta$, and $\gamma$ encode model-dependent deviations
from Schwarzschild gravity inspired from [Phys.Rev.D 112 (2025) 12, 124072]. Starting from the exact first-order
orbit equation, we derive a compact master equation for the impact-parameter-dependent
trajectory $u(\varphi)\equiv 1/r(\varphi)$ and obtain a model-independent
expression for the bending angle $\alpha(b)$ in terms of generic metric
functions and their derivatives. This master equation is then solved
semi-analytically by three complementary techniques: (i) the homotopy
perturbation method (HPM), (ii) the variational iteration method (VIM),
and (iii) a calibrated impulse (single-kick) approximation expressed
directly in terms of the effective gravitational potential.
As nontrivial test beds we consider a scalar-hairy Reissner-Nordstr\"om-like
black hole where the scalar hair enters as $Q_s$ in an effective
charge parameter. Then we derive closed-form
expressions for the deflection angle, identify the leading scalar-hair, and compare the accuracy and convergence
properties of HPM, VIM, and the impulse method against the standard
Schwarzschild limits. Our results show that the
generic formulation in $(\alpha,\beta,\gamma)$ can efficiently accommodate
a broad class of modified gravity black hole solutions. \textcolor{black}{We further supplement the analytic treatment with a compact numerical results against the exact null-geodesic integral near the photon sphere in order to delineate the practical range of validity of the three approximation schemes.}
\end{abstract}


\date{\today}

\maketitle

%\tableofcontents

\section{Introduction}
The gravitational deflection of light stands as a foundational prediction of Einstein's theory of General Relativity (GR), transitioning from a crucial early test of the theory \cite{Weinberg:1972kfs, Will:2014kxa} to a vital observational tool in modern physics and astrophysics \cite{Schneider:1992bmb, Narayan:1996ba}. While the majority of observed gravitational lensing phenomena occur in the weak-field limit where the deflection angle is small \cite{Virbhadra:1999nm,Bozza:2010xqn}, the extreme spacetime curvature near compact objects such as black holes provides a unique arena for testing GR in the strong-deflection regime \cite{Virbhadra:2008ws,Bozza:2001xd}. In this regime, light rays can undergo large deflections, potentially completing one or more orbits around the lens before reaching an observer, giving rise to a theoretically infinite sequence of highly demagnified "relativistic images" \cite{Virbhadra:2002ju,Claudel:2000yi,Virbhadra:1998dy,Virbhadra:2007kw,Virbhadra:2022iiy,Virbhadra:2024xpk}.

Calculating the deflection angle, which is the key observable in lensing phenomena, presents significant analytical challenges. The exact null geodesic equations often lead to solutions involving elliptic integrals, which are computationally intensive and can obscure physical insight \cite{Shchigolev:2016gro, Keeton:2005jd,Li:2019mqw,Huang:2023bto,Huang:2025vqm}. This has motivated the development of various approximative and semi-analytical methods to obtain accurate solutions.  The formalism for testing alternative theories of gravity via gravitational lensing by static, spherically symmetric compact objects was systematically developed in \cite{Keeton:2005jd}. The application of the Gauss-Bonnet theorem to gravitational lensing, offering a novel geometric perspective, was first introduced by Gibbons and Werner \cite{Gibbons:2008rj,Gibbons:2008hb}. Ishihara et al. extended the Gauss-Bonnet approach to calculate the bending angle of light for observers and sources at finite distances from the lens \cite{Ishihara:2016vdc}. Werner formulated the theory of gravitational lensing in the Kerr-Randers optical geometry, enabling the study of lensing in rotating spacetimes \cite{Werner:2012rc}. Then authors investigated the finite-distance gravitational deflection of massive particles in a Kerr-like black hole spacetime within the bumblebee gravity framework \cite{Li:2020dln}.

Standard approaches often rely on a post-Newtonian (PPN) expansion, which is well-suited for the weak-field limit but can be extended to higher orders to capture more subtle relativistic effects \cite{Keeton:2005jd, Iyer:2009wa}. For deflections near the photon sphere, the Strong Deflection Limit (SDL) provides a robust logarithmic approximation \cite{Bozza:2001xd}. Extensive progress in the study of black hole shadows, weak gravitational lensing, plasma effects, and particle dynamics across various modified gravity theories has been achieved through numerous investigations \cite{Atamurotov:2015nra,Atamurotov:2013sca,Mustafa:2022xod,Abdujabbarov:2017pfw,Atamurotov:2021imh,Atamurotov:2022knb,Atamurotov:2021qds,Atamurotov:2022slw}. A broad range of recent advances in black hole physics, plasma lensing, nonlinear electrodynamics, and modified gravity theories has been driven by detailed studies of particle dynamics, photon motion, oscillatory phenomena, and thermodynamical analyses across diverse gravitational backgrounds \cite{Rahmatov:2025gpk,Rayimbaev:2023bjs,Rayimbaev:2021vsq,Rayimbaev:2021luv,Ditta:2023wye,Ditta:2023ccf,Ditta:2023rhr,Mustafa:2024zsx,Ashraf:2024dwg,Feng:2024jeq,Ashraf:2025cnz,Mushtaq:2025ewk,Ditta:2025bri}.

Beyond the standard framework of GR, a variety of alternative theories of gravity have been proposed to address open questions in cosmology, such as the nature of dark energy, or to explore the consequences of fundamental symmetry breaking. Spherically symmetric solutions describing black holes surrounded by quintessential fields, for example, have been explored as models incorporating dark energy \cite{Shchigolev:2016gro}. Gravitational lensing by black holes in these modified theories provides a promising observational window to constrain their parameters and test their validity against standard GR \cite{Keeton:2005jd}.

The equation of motion for a photon in a static, spherically symmetric spacetime is typically a second-order nonlinear ordinary differential equation. For the simplest case of the Schwarzschild metric, the equation for the path $u(\varphi)$, where $u = 1/r$, is given by:
\begin{equation} \label{eq:orbit}
    \frac{d^2u}{d\varphi^2} + u = 3mu^2
\end{equation}
where $m$ is the mass of the lensing object in geometrized units. While this equation can be solved exactly in terms of elliptic integrals, such solutions are often cumbersome. For more complex spacetimes, exact analytic solutions are generally not
available in compact elementary form and may obscure the dependence on
the physical parameters.

The complexity of the geodesic equations in these alternative spacetimes necessitates the use of efficient and reliable computational tools. Among these, semi-analytical techniques such as the Homotopy-Perturbation Method (HPM) and the Variational Iteration Method (VIM) have proven to be particularly effective \cite{he1999, he2000}. These methods construct an approximate solution in the form of a rapidly converging series without inserting an explicit perturbation parameter into the field equations; in lensing applications the relevant smallness enters through weak-field boundary terms and controls truncation accuracy \cite{Shchigolev:2015sgg, Shchigolev:2016vpz, Shchigolev:2016gro}. Complementing these is the Single-Impulse, or Transverse-Kick, Method, which provides a physically intuitive picture by approximating the gravitational deflection as a discrete impulse perpendicular to the photon's path.

\textcolor{black}{From a methodological perspective, the three approaches used in this work are complementary rather than redundant. In HPM one embeds the full nonlinear orbit equation into a one-parameter homotopy that continuously deforms the exactly solvable straight-line problem into the curved-spacetime geodesic problem; the resulting hierarchy of linear equations yields systematic corrections order by order in the weak-field quantities \cite{he1999,he2000,Shchigolev:2016gro}. In VIM one constructs a correction functional with an optimally determined Lagrange multiplier, so that each iterate improves the previous trajectory while preserving the functional form of the solution and avoiding discretization \cite{he1999,Shchigolev:2016vpz,HE20073}. The impulse approximation, by contrast, treats the bending as the accumulated transverse momentum kick generated by an effective potential along an almost straight ray, making the physical origin of the deflection especially transparent and yielding very fast closed-form estimates in parameter scans \cite{Weinberg:1972kfs,Misner:1973prb,Schutz:1985jx}. The practical advantage of combining these methods is therefore clear: HPM and VIM provide systematic semi-analytic control over the nonlinear geodesic equation, whereas the impulse method offers intuition and speed. Because all three are weak-field constructions, their domain of reliability must still be checked against the exact geodesic solution as one approaches the photon sphere, where the bending angle grows rapidly. For this reason, besides deriving the analytic formulas, below we also include a dedicated numerical results near the photon sphere.}

Most semi-analytical treatments of weak gravitational lensing are tailored
to specific metrics, such as the Schwarzschild or Reissner-Nordstr\"om
solutions, and must be re-derived from scratch whenever one considers
hairy or modified gravity black holes. In view of the rapidly growing
landscape of viable black hole solutions, it is desirable to construct
a framework that starts from a generic static, spherically symmetric
line element
\(
ds^2 = -\alpha(r,\delta)\,dt^2 + \gamma(r,\delta)\,dr^2 +
\beta(r,\delta)\,d\Omega_2^2,
\)
and only afterwards specializes to particular models.
In this work we show that the homotopy perturbation method, the variational
iteration method, and the impulse (single-kick) approximation can all be
formulated directly at this generic level. The model dependence then
enters only through the functions $\alpha$, $\beta$, and $\gamma$,
allowing one to treat scalar hair \cite{Astorino:2013sfa}  within a single semi-analytical
lensing toolbox.

For a generic static spherically symmetric metric
\(
ds^2=-\alpha(r)\,dt^2+\gamma(r)\,dr^2+\beta(r)\,d\Omega^2,
\)
null geodesics confined to the equatorial plane admit the standard first integral
\(
\gamma(r)\dot r^2=\frac{E^2}{\alpha(r)}-\frac{L^2}{\beta(r)}.
\)
With impact parameter \(b=L/E\) and closest approach \(r_0\) determined by
\(b^2=\beta(r_0)/\alpha(r_0)\),
the total bending angle can be written exactly as the one-dimensional integral
\begin{equation}
\hat\alpha(b)=2\int_{r_0}^{\infty}
\frac{b\,\sqrt{\alpha(r)\gamma(r)}}{\sqrt{\beta(r)}\sqrt{\beta(r)-\alpha(r)b^2}}\;dr-\pi.
\label{eq:exact_bending_generic}
\end{equation}

\textcolor{black}{Equation~\eqref{eq:exact_bending_generic} is exact for null rays
with a real outer turning point \(r_0\) satisfying
\(b^2=\beta(r_0)/\alpha(r_0)\), provided the trajectory lies outside
the critical photon orbit so that
\(\beta(r)-\alpha(r)b^2>0\) for \(r>r_0\). In asymptotically flat
spacetimes the subtraction by \(\pi\) gives the usual total bending
angle; in asymptotically conical or locally flat geometries this
subtraction must be modified by the corresponding asymptotic opening
angle.}

The motivation for developing weak-field approximation schemes is therefore not
the absence of an exact integral representation, but (i) the need for explicit
closed-form coefficient dependence on model parameters for lens modeling and
parameter inference, (ii) efficient evaluation in large parameter scans, and
(iii) applicability when the metric functions are known only numerically or
as asymptotic series, where an analytic PM/PPN-type expansion is the natural
observable-oriented output.

\section{Homotopy Perturbation Method for Generic static spherically symmetric spacetime}
\label{sec:generic-HPM}

Many of the fundamental equations in physics, particularly the geodesic equations governing the paths of light rays in General Relativity, are inherently nonlinear \cite{Shchigolev:2016vpz, Shchigolev:2016gro}. This nonlinearity often precludes exact analytical solutions, necessitating the use of numerical or approximative techniques \cite{Shchigolev:2016vpz}. While traditional perturbation methods are powerful, they typically depend on the existence of a small physical parameter, which is not always available or convenient \cite{Shchigolev:2016gro,Keeton:2005jd}.

The Homotopy Perturbation Method (HPM), first proposed by Ji-Huan He, is a potent semi-analytical technique developed to overcome these limitations when solving nonlinear differential and integral equations \cite{he1999,he2000, Shchigolev:2016gro}. The core of the method is the construction of a homotopy (a continuous deformation) that links a simple, solvable problem to the complex, original nonlinear problem. A solution is then sought as a power series of an embedding parameter $p \in [0, 1]$, which deforms the simple problem (where $p=0$) into the original one (where $p=1$).

The motivation for employing HPM in gravitational physics is compelling due to its distinct advantages.Although the HPM embedding parameter is auxiliary, the accuracy of a
truncated lensing solution is still controlled by weak-field ratios such
as \(m/b\), \(q^{2}/b^{2}\), and the relevant deformation scales \cite{Shchigolev:2016gro}. Furthermore, the method avoids discretization and linearization, preserving the full nonlinear structure of the problem. HPM typically yields a solution in the form of a rapidly converging infinite series, where the first few terms alone often provide a highly accurate approximation of the exact solution. These features have led to its successful application across numerous fields of science and, more recently, to specific problems in cosmology and astrophysics, such as calculating the perihelion precession of Mercury and the gravitational deflection of light.

We now extend the above construction to a generic static, spherically
symmetric spacetime of the form \cite{Pantig:2025deu}
\begin{equation}
ds^2 = -\alpha(r,\delta)\,dt^2
      + \gamma(r,\delta)\,dr^2
      + \beta(r,\delta)\,d\Omega_2^2,
\label{eq:generic_metric}
\end{equation}
where $d\Omega_2^2 = d\theta^2 + \sin^2\theta\,d\phi^2$, and
$\alpha,\gamma,\beta$ are arbitrary positive functions of $r$ and of a
deformation parameter $\delta$. In the Minkowski limit one has $\alpha\to 1$, $\gamma\to 1$, and $\beta\to r^2$, so that
Eq.~\eqref{eq:generic_metric} reduces to flat spacetime in spherical coordinates.

Restricting to the equatorial plane $\theta=\pi/2$, the metric reads
\begin{equation}
ds^2 = -\alpha(r,\delta)\,dt^2
      + \gamma(r,\delta)\,dr^2
      + \beta(r,\delta)\,d\phi^2.
\label{eq:generic_equatorial}
\end{equation}
The Killing vectors $\partial_t$ and $\partial_\phi$ provide the conserved
energy $E$ and angular momentum $L$:
\begin{equation}
E = \alpha(r,\delta)\,\frac{dt}{d\lambda},\qquad
L = \beta(r,\delta)\,\frac{d\phi}{d\lambda},
\end{equation}
where $\lambda$ is an affine parameter.  The null condition $ds^2=0$ gives
\begin{equation}
0 = -\alpha\left(\frac{dt}{d\lambda}\right)^2
    + \gamma\left(\frac{dr}{d\lambda}\right)^2
    + \beta\left(\frac{d\phi}{d\lambda}\right)^2,
\end{equation}
so that
\begin{equation}
\gamma(r,\delta)\left(\frac{dr}{d\lambda}\right)^2
= \frac{E^2}{\alpha(r,\delta)} - \frac{L^2}{\beta(r,\delta)}.
\label{eq:generic_radial_first}
\end{equation}

Introduce the impact parameter $b \equiv L/E$ and the inverse radius
\begin{equation}
u(\phi) \equiv \frac{1}{r(\phi)}.
\end{equation}

Then
\begin{equation}
\frac{dr}{d\phi}
= \frac{dr/d\lambda}{d\phi/d\lambda}
= \frac{\beta(r,\delta)}{L}\frac{dr}{d\lambda},
\end{equation}
and Eq.~\eqref{eq:generic_radial_first} yields
\begin{equation}
\left(\frac{dr}{d\phi}\right)^2
= \frac{\beta^2(r,\delta)}{L^2}\frac{1}{\gamma(r,\delta)}
\left(\frac{E^2}{\alpha(r,\delta)} - \frac{L^2}{\beta(r,\delta)}\right).
\end{equation}
In terms of $u=1/r$ we have
\begin{equation}
\frac{dr}{d\phi} = -\frac{1}{u^2}\frac{du}{d\phi},
\qquad
\left(\frac{du}{d\phi}\right)^2
= u^4 \left(\frac{dr}{d\phi}\right)^2.
\end{equation}
Using $L=bE$ and writing the metric functions as functions of $u$ via
$r=1/u$, we obtain
\begin{equation}
\left(\frac{du}{d\phi}\right)^2
= \frac{u^4}{\gamma(u,\delta)\,b^2}
\left(\frac{\beta^2(u,\delta)}{\alpha(u,\delta)}
      - b^2\,\beta(u,\delta)\right).
\label{eq:uprime_sq_generic}
\end{equation}
For later convenience, define the function
\begin{equation}
\mathcal{S}(u,\delta) \equiv
\frac{u^4}{\gamma(u,\delta)}
\left(\frac{\beta^2(u,\delta)}{\alpha(u,\delta)}
      - b^2\,\beta(u,\delta)\right).
\label{eq:S_def}
\end{equation}
Then Eq.~\eqref{eq:uprime_sq_generic} can be written compactly as
\begin{equation}
\left(\frac{du}{d\phi}\right)^2
= \frac{1}{b^2}\,\mathcal{S}(u,\delta).
\end{equation}
Differentiating with respect to $\phi$ yields
\begin{equation}
2u' u'' = \frac{1}{b^2}\,\frac{\partial \mathcal{S}}{\partial u}\,u',
\end{equation}
so for generic points with $u'\neq 0$ we obtain the second-order null
geodesic equation
\begin{equation}
u'' = \frac{1}{2b^2}\,\frac{\partial \mathcal{S}(u,\delta)}{\partial u}.
\label{eq:uPP_generic_S}
\end{equation}

In flat spacetime, $\alpha=\gamma=1$, $\beta=r^2=1/u^2$, and one finds
$\mathcal{S}_{\rm flat}(u) = 1 - b^2 u^2$, so that
Eq.~\eqref{eq:uPP_generic_S} reduces to $u'' = -u$, i.e.
\begin{equation}
u'' + u = 0 \qquad (\text{Minkowski}).
\end{equation}
This allows us to write the general equation as a deformation of the flat
case:
\begin{equation}
u'' + u = \mathcal{N}(u,\delta),
\label{eq:generic_geodesic_N}
\end{equation}
where the \emph{nonlinear source} $\mathcal{N}$ encodes the deviation of the
metric from Minkowski:
\begin{equation}
\mathcal{N}(u,\delta) \equiv
\frac{1}{2b^2}\,\frac{\partial \mathcal{S}(u,\delta)}{\partial u} + u.
\label{eq:N_def}
\end{equation}
By construction $\mathcal{N}(u,\delta)\to 0$ in the flat limit.

We now apply the homotopy perturbation method to
Eq.~\eqref{eq:generic_geodesic_N}.  We assume that the metric functions
$\alpha,\gamma,\beta$ can be expanded around a reference (e.g.\ Minkowski
or Schwarzschild) in terms of small parameters $\epsilon_i$ (which may
include the mass parameter $\epsilon_1=m/b$ and the deformation parameter
$\epsilon_2=\delta$), so that
\begin{equation}
\mathcal{N}(u,\delta) = \mathcal{O}(\epsilon_1,\epsilon_2,\ldots).
\end{equation}

Introduce an embedding parameter $p\in[0,1]$ and construct the homotopy
\begin{equation}
u'' + u - p\,\mathcal{N}(u,\delta) = 0.
\label{eq:homotopy_generic}
\end{equation}
For $p=0$ we recover the straight-line equation $u''+u=0$, and for $p=1$
we recover the full geodesic equation~\eqref{eq:generic_geodesic_N}.

We seek a series solution in $p$,
\begin{equation}
u(\phi) = u_0(\phi) + p\,u_1(\phi) + p^2 u_2(\phi) + \cdots,
\label{eq:u_series_generic}
\end{equation}
with boundary conditions chosen such that the zeroth-order solution
describes a straight line with impact parameter $b$:
\begin{equation}
u_0(0)=0,\quad u_0'(0)=\frac{1}{b},\qquad
u_n(0)=u_n'(0)=0\quad (n\geq 1).
\label{eq:ICs_generic}
\end{equation}
Substituting~\eqref{eq:u_series_generic} into
Eq.~\eqref{eq:homotopy_generic} and collecting powers of $p$ gives the
hierarchy
\begin{align}
\mathcal{O}(p^0):\quad & u_0'' + u_0 = 0, 
\label{eq:u0_generic}\\
\mathcal{O}(p^1):\quad & u_1'' + u_1 = \mathcal{N}(u_0,\delta), 
\label{eq:u1_generic}\\
\mathcal{O}(p^2):\quad & u_2'' + u_2 = 
\mathcal{N}'(u_0,\delta)\,u_1,\quad\ldots,
\label{eq:u2_generic}
\end{align}
where $\mathcal{N}'(u_0,\delta) \equiv \partial \mathcal{N}/\partial u$
evaluated at $u_0$.

The zeroth-order solution is simply
\begin{equation}
u_0(\phi) = \frac{1}{b}\sin\phi,
\end{equation}
and the first-order correction satisfies the linear inhomogeneous equation
\begin{equation}
u_1'' + u_1 = \mathcal{N}\!\left(u_0(\phi),\delta\right),
\label{eq:u1_source_generic}
\end{equation}
with $u_1(0)=u_1'(0)=0$.  The source term on the right-hand side is
completely determined by the metric functions via
Eqs.~\eqref{eq:S_def}--\eqref{eq:N_def}, evaluated along the straight-line
trajectory $u_0(\phi)=(1/b)\sin\phi$.

In practice, one first expands $\mathcal{N}(u,\delta)$ in small parameters,
\begin{equation}
\mathcal{N}(u,\delta)
= \sum_i \epsilon_i\,\mathcal{N}_i(u)
+ \sum_{ij}\epsilon_i\epsilon_j\,\mathcal{N}_{ij}(u)
+ \cdots,
\end{equation}
and then solves Eq.~\eqref{eq:u1_source_generic} order by order in
$\epsilon_i$.  The second-order term $u_2(\phi)$ in
Eq.~\eqref{eq:u2_generic} similarly gives access to the
$\mathcal{O}(\epsilon_i^2)$ corrections to the trajectory.

Once the approximate trajectory $u(\phi)$ is known, the weak deflection
angle can be extracted using the same root-equation method employed in
Sec.~\ref{sec:generic-HPM}.  We write the solution as
\begin{equation}
u(\phi) = \frac{1}{b}\sin\phi + \frac{1}{b}U(\phi),
\end{equation}
where $U(\phi)$ collects all corrections due to the curvature and the
deformation parameter~$\delta$ and satisfies $|U(\phi)|\ll 1$ in the weak
field regime.

The condition for the outgoing ray to reach infinity is
\begin{equation}
u(\phi)=0
\quad\Rightarrow\quad
\sin\phi + U(\phi) = 0.
\end{equation}
Let $\phi_\infty = \pi + \beta$ denote the outgoing asymptotic direction,
where $\beta$ is the total deflection angle.  Following the homotopy
construction for the root equation (see also Ref.~\cite{Shchigolev:2016gro}),
one introduces
\begin{equation}
\sin\phi + q\,U(\phi) = 0,\qquad q\in[0,1],
\end{equation}
and expands
\begin{equation}
\phi = \phi_0 + q\,\phi_1 + q^2 \phi_2 + \cdots.
\end{equation}
Taking $\phi_0=\pi$ and matching powers of $q$ yields algebraic equations
for $\phi_1,\phi_2,\ldots$ in terms of $U(\pi)$ and its derivatives.
Up to third order one finds the general expression

Expanding
\begin{equation}
\sin(\pi+\beta)+U(\pi+\beta)=0
\end{equation}
for \(|\beta|\ll1\), one obtains
\begin{equation}
\beta \simeq U(\pi)
\left[
1+U'(\pi)+U'^2(\pi)
+\frac{1}{2}U(\pi)U''(\pi)
+\frac{1}{6}U^2(\pi)
\right].
\label{eq:root_expansion_corrected}
\end{equation}

where $U'(\pi)$ and $U''(\pi)$ are derivatives with respect to $\phi$.
In the generic spacetime~\eqref{eq:generic_metric}, the functions
$U(\pi)$, $U'(\pi)$ and $U''(\pi)$ are determined by the HPM solution
$u(\phi)$ and ultimately by the metric functions $\alpha,\gamma,\beta$ and
the deformation parameter $\delta$ via $\mathcal{N}(u,\delta)$.

Thus, the homotopy perturbation method provides a systematic and universal
framework to compute the photon trajectory and the weak deflection angle
for any static, spherically symmetric spacetime of the form
Eq.~\eqref{eq:generic_metric}, treating deviations from Minkowski or
Schwarzschild as small perturbations encoded in the source
$\mathcal{N}(u,\delta)$.

\section{Variational iteration method for a generic static, spherically symmetric spacetime}
\label{sec:vim-generic}

Among the spectrum of analytical approximation techniques, the Variational Iteration Method (VIM) has emerged as a robust and versatile tool for solving nonlinear differential equations. First introduced by He~\cite{he1999}, VIM circumvents the need for small expansion parameters, linearization, or discretization, which are features that are particularly advantageous in General Relativity, where perturbative approaches often suffer from slow convergence or limited applicability.

The motivation for employing VIM in calculating the gravitational deflection of light lies in its inherent strengths in handling nonlinearities. Traditional perturbative techniques typically rely on the existence of a small parameter, such as the ratio of the Schwarzschild radius to the impact parameter, which may not always be justified or sufficiently small. In contrast, VIM operates without such assumptions, rendering it applicable to a wider array of physical regimes, including those characterized by strong gravitational fields.

The core of the VIM approach is the construction of a correction functional for the differential equation in question~\cite{Shchigolev:2016vpz}. This functional incorporates a general Lagrange multiplier, optimally determined via variational principles, which generates a recursive sequence of approximations. Notably, the nonlinear terms are handled as restricted variations ($\delta\tilde{u}_n = 0$), simplifying the identification of the Lagrange multiplier and facilitating a straightforward iterative scheme. Starting from an initial guess-often the solution to the corresponding linearized problem-VIM yields a rapidly convergent sequence of approximations, frequently requiring only a few iterations to achieve high accuracy~\cite{HE20073}. Moreover, the method retains the solution in a continuous, functional form, preserving physical transparency and interpretability, in contrast to discretization-based approaches. Its simple iterative structure and the straightforward determination of the Lagrange multiplier have enabled successful applications to a variety of both ordinary and partial differential equations~\cite{WAZWAZ2007895}.

Within the context of gravitational lensing, the principal advantages of VIM are particularly pronounced. Like the Homotopy Perturbation Method (HPM), VIM does not depend on the presence of small expansion parameters, enhancing its applicability to problems where such parameters are either absent or not sufficiently small for standard perturbation theory to be reliable~\cite{Shchigolev:2016vpz}. Each VIM iteration typically reduces to a simple integration, making the method computationally efficient and easy to implement. 

In this section we generalize the variational iteration method (VIM) approach
for orbital motion and light deflection to a generic static, spherically
symmetric spacetime, described by the line element
\begin{equation}
    ds^2 = -\alpha(r,\delta)\,dt^2 
           + \gamma(r,\delta)\,dr^2 
           + \beta(r,\delta)\,d\Omega_2^2,
    \label{eq:metric-generic-r}
\end{equation}
where
\begin{equation}
    d\Omega_2^2 = d\theta^2 + \sin^2\theta\,d\varphi^2,
\end{equation}
and $\delta$ denotes a deformation parameter. We assume
$\beta(r,\delta)>0$ and spherical symmetry, and we restrict the motion to the
equatorial plane $\theta=\pi/2$ without loss of generality.

It is convenient to introduce the \emph{areal radius}
\begin{equation}
    R \equiv \sqrt{\beta(r,\delta)} ,
    \label{eq:areal-radius-def}
\end{equation}
which ensures that the area of the symmetry two-spheres is $4\pi R^2$. We assume
that, for fixed $\delta$, the map $r\mapsto R(r,\delta)$ is monotonic and thus
invertible, so that we may write $r=r(R,\delta)$.

In the coordinates $(t,R,\theta,\varphi)$, the angular part of
Eq.~\eqref{eq:metric-generic-r} takes the standard form
\begin{equation}
    \beta(r,\delta)\,d\Omega_2^2 = R^2\,d\Omega_2^2 .
\end{equation}
The radial part transforms as
\begin{equation}
    \gamma(r,\delta)\,dr^2 
    = \gamma(r,\delta)\left(\frac{dr}{dR}\right)^2 dR^2
    \equiv \frac{dR^2}{H(R,\delta)},
\end{equation}
which defines the function
\begin{equation}
    H(R,\delta)
    \equiv \frac{1}{
    \gamma\!\big(r(R,\delta),\delta\big)\,
    \left(\dfrac{dr}{dR}\right)^2} .
    \label{eq:H-def}
\end{equation}
Similarly, we define
\begin{equation}
    A(R,\delta) \equiv 
    \alpha\!\big(r(R,\delta),\delta\big) .
    \label{eq:A-def}
\end{equation}

With these definitions the metric \eqref{eq:metric-generic-r} is cast into the
canonical static, spherically symmetric form
\begin{equation}
    ds^2 = -A(R,\delta)\,dt^2 
           + \frac{dR^2}{H(R,\delta)} 
           + R^2\,d\Omega_2^2.
    \label{eq:metric-canonical}
\end{equation}
The standard Schwarzschild-type coordinates correspond to the special choice
$A(R,\delta)=H(R,\delta)=1-2M/R$; more general modified gravity or
nonlinear-electrodynamics spacetimes are encoded in the functional forms of
$A$ and $H$.

Restricting to the equatorial plane $\theta=\pi/2$, the line element
\eqref{eq:metric-canonical} reduces to
\begin{equation}
    ds^2 = -A(R)\,dt^2 
           + \frac{dR^2}{H(R)} 
           + R^2 d\varphi^2,
    \label{eq:metric-equatorial}
\end{equation}
where, for brevity, we suppress the explicit dependence on $\delta$ in
$A(R,\delta)$ and $H(R,\delta)$.

We consider geodesics parametrized by an affine parameter $\tau$, with the
Lagrangian
\begin{equation}
    2\mathcal{L} =
    -A(R)\,\dot t^{\,2} 
    + \frac{1}{H(R)}\,\dot R^{\,2} 
    + R^2 \dot\varphi^{\,2}
    = -k,
    \label{eq:lagrangian}
\end{equation}
where an overdot denotes $d/d\tau$, and
\begin{equation}
    k=
    \begin{cases}
        1, & \text{timelike geodesics (massive test particles)}, \\
        0, & \text{null geodesics (light rays)} .
    \end{cases}
\end{equation}

The metric is static and spherically symmetric, so the Killing vectors
$\partial_t$ and $\partial_\varphi$ generate two conserved quantities:
\begin{equation}
    E = A(R)\,\dot t,
    \qquad
    L = R^2 \dot\varphi,
    \label{eq:EL-def}
\end{equation}
which may be interpreted as the specific energy and angular momentum of the
particle.

Using Eqs.~\eqref{eq:lagrangian} and \eqref{eq:EL-def}, the normalization
condition for the four-velocity becomes
\begin{equation}
    -\frac{E^2}{A(R)} 
    + \frac{1}{H(R)}\,\dot R^{\,2} 
    + \frac{L^2}{R^2}
    = -k,
\end{equation}
which can be rearranged to yield the radial equation
\begin{equation}
    \dot R^{\,2} 
    = H(R)\left[\frac{E^2}{A(R)} - \frac{L^2}{R^2} - k\right].
    \label{eq:R-dot}
\end{equation}

To obtain the orbit equation we switch from $\tau$ to $\varphi$ as the
independent variable. From Eq.~\eqref{eq:EL-def} we have
\begin{equation}
    \frac{dR}{d\varphi}
    = \frac{\dot R}{\dot\varphi}
    = \frac{\dot R}{L/R^2}
    = \frac{R^2}{L}\,\dot R,
\end{equation}
and therefore
\begin{equation}
    \left(\frac{dR}{d\varphi}\right)^2
    = \frac{R^4}{L^2}\,\dot R^{\,2}
    = \frac{R^4}{L^2}\,
      H(R)\left[\frac{E^2}{A(R)} - \frac{L^2}{R^2} - k\right].
    \label{eq:R-phi-eq}
\end{equation}

Following standard practice, we introduce the inverse radius
\begin{equation}
    u(\varphi) \equiv \frac{1}{R(\varphi)} .
    \label{eq:u-def}
\end{equation}
Then
\begin{equation}
    \frac{dR}{d\varphi} = -\frac{1}{u^2}\,\frac{du}{d\varphi},
    \qquad
    \left(\frac{dR}{d\varphi}\right)^2
    = \frac{1}{u^4}\left(\frac{du}{d\varphi}\right)^2.
\end{equation}
Using $R=1/u$ in Eq.~\eqref{eq:R-phi-eq} we obtain
\begin{equation}
    \frac{1}{u^4}\left(\frac{du}{d\varphi}\right)^2
    = \frac{(1/u)^4}{L^2}\,H(u)
      \left[\frac{E^2}{A(u)} - L^2 u^2 - k\right],
\end{equation}
where we have written $A(u)\equiv A(R=1/u)$, $H(u)\equiv H(R=1/u)$. The factors
of $u^4$ cancel, and we are left with the first-order equation
\begin{equation}
    \left(\frac{du}{d\varphi}\right)^2
    = \frac{H(u)}{A(u)}\left(\frac{E^2}{L^2}\right)
      - H(u)\left(u^2 + \frac{k}{L^2}\right).
    \label{eq:du-dphi-sq}
\end{equation}

Differentiating Eq.~\eqref{eq:du-dphi-sq} with respect to $\varphi$, and using
$u' \equiv du/d\varphi$, $u'' \equiv d^2u/d\varphi^2$, we note that
\begin{equation}
    u'' = \frac{1}{2}\frac{d}{du}\big(u'^2\big)
         = \frac{1}{2}\frac{d}{du}
         \left[
           \frac{H(u)}{A(u)}\left(\frac{E^2}{L^2}\right)
           - H(u)\left(u^2 + \frac{k}{L^2}\right)
         \right].
\end{equation}
A straightforward differentiation yields the \emph{master orbit equation}
\begin{equation}
    \frac{d^2u}{d\varphi^2}
    = \frac{E^2}{2L^2}\frac{d}{du}\!
      \left(\frac{H(u)}{A(u)}\right)
      - H(u)\,u
      - \frac{1}{2}\left(u^2 + \frac{k}{L^2}\right)\frac{dH(u)}{du}.
    \label{eq:master-orbit}
\end{equation}
This is the generalization of the familiar orbit equation in Schwarzschild
spacetime to an arbitrary static, spherically symmetric geometry specified by
the functions $A(u,\delta)$ and $H(u,\delta)$.

For the application of He's variational iteration method it is convenient to
bring Eq.~\eqref{eq:master-orbit} into the form
\begin{equation}
    \frac{d^2u}{d\varphi^2} + u = \mathcal{S}(u;\delta),
    \label{eq:orbit-vim-form}
\end{equation}
where $\mathcal{S}(u;\delta)$ collects all nonlinear and deformation-dependent
terms. Adding $u$ to both sides of Eq.~\eqref{eq:master-orbit} we obtain
\begin{equation}
    \frac{d^2u}{d\varphi^2} + u
    = \frac{E^2}{2L^2}\frac{d}{du}\!
      \left(\frac{H}{A}\right)
      - H(u)\,u
      - \frac{1}{2}\left(u^2 + \frac{k}{L^2}\right)\frac{dH}{du}
      + u .
\end{equation}
We therefore define
\begin{equation}
    \mathcal{S}(u;\delta)
    \equiv
    \frac{E^2}{2L^2}\frac{d}{du}\!
      \left(\frac{H(u,\delta)}{A(u,\delta)}\right)
      - H(u,\delta)\,u
      - \frac{1}{2}\left(u^2 + \frac{k}{L^2}\right)
        \frac{\partial H(u,\delta)}{\partial u}
      + u.
    \label{eq:S-def}
\end{equation}
In the Schwarzschild limit $A=H=1-2Mu$ one recovers the familiar equation
$u''+u = 3Mu^2 + kM/L^2$, while any modified gravity or nonlinear
electrodynamics contribution is encapsulated in the functional dependence of
$\mathcal{S}$ on $u$ and the parameter $\delta$.

Following He's variational iteration method, we rewrite
Eq.~\eqref{eq:orbit-vim-form} in operator form
\begin{equation}
    L[u] + N[u] = 0,
\end{equation}
where
\begin{equation}
    L[u] \equiv \frac{d^2u}{d\varphi^2} + u,
    \qquad
    N[u] \equiv -\mathcal{S}(u;\delta).
\end{equation}
The corresponding correction functional is chosen as
\begin{equation}
    u_{n+1}(\varphi)
    = u_{n}(\varphi)
    + \int_{0}^{\varphi} \lambda(\phi,\varphi)
      \bigg[
        u_{n}''(\phi) + u_{n}(\phi)
        - \mathcal{S}\big(\tilde u_{n}(\phi);\delta\big)
      \bigg] d\phi,
    \label{eq:VIM-functional}
\end{equation}
where $\tilde u_{n}$ denotes a restricted variation, $\delta\tilde u_n=0$, and
$\lambda(\phi,\varphi)$ is a Lagrange multiplier to be determined.

Requiring stationarity of the correction functional with respect to arbitrary
variations $\delta u_{n}$ yields the differential equation
\begin{equation}
    \frac{\partial^2\lambda}{\partial\phi^2}
    + \lambda = 0
\end{equation}
together with the conditions
\begin{equation}
    \lambda(\phi,\varphi)\big|_{\phi=\varphi} = 0,
    \qquad
    \frac{\partial\lambda}{\partial\phi}(\phi,\varphi)\big|_{\phi=\varphi}
    = 1.
\end{equation}
Solving this boundary-value problem leads to
\begin{equation}
    \lambda(\phi,\varphi) = \sin(\phi-\varphi).
\end{equation}
Importantly, this Lagrange multiplier depends only on the choice of linear
operator $L[u]=u''+u$ and is \emph{independent} of the specific form of
$\mathcal{S}(u;\delta)$. As a result, the VIM iteration scheme retains the
same structure for any static, spherically symmetric spacetime.

Substituting $\lambda(\phi,\varphi)=\sin(\phi-\varphi)$ into
Eq.~\eqref{eq:VIM-functional}, we obtain the generic iteration formula
\begin{equation}
u_{n+1}(\varphi)
    = u_{n}(\varphi)
    + \int_{0}^{\varphi} \sin(\phi-\varphi)\,
      \bigg[
        u_{n}''(\phi) + u_{n}(\phi)
        - \mathcal{S}\big(u_{n}(\phi);\delta\big)
      \bigg] d\phi.
    \label{eq:VIM-iter-general}
\end{equation}
Given an appropriate zeroth-order approximation $u_0(\varphi)$, this recurrence
relation yields increasingly accurate approximations to the exact orbit
$u(\varphi)$.

To extract physical observables such as perihelion precession for massive
particles and deflection angles for light rays, suitable initial approximations
are required.

For \emph{timelike} geodesics ($k=1$) corresponding to bound orbits, we adopt
a Keplerian-type ansatz,
\begin{equation}
    u_0(\varphi)
    = u_c\big(1 + e\cos\varphi\big),
    \label{eq:u0-timelike}
\end{equation}
where $e$ is the orbital eccentricity and $u_c$ is the inverse radius of the
associated circular orbit, determined by the condition of extremum of the
effective potential. Substituting Eq.~\eqref{eq:u0-timelike} into
\eqref{eq:VIM-iter-general} yields the first iterate $u_1(\varphi)$, from which
the perihelion advance $\Delta\varphi$ can be extracted by imposing
$du/d\varphi=0$ at perihelion and writing the azimuthal coordinate as
$\varphi=2\pi+\Delta\varphi$, with $\Delta\varphi\ll1$.

For \emph{null} geodesics ($k=0$), describing light rays with impact parameter
$b$, the zeroth-order trajectory in the weak-field limit is taken to be a
straight line,
\begin{equation}
    u_0(\varphi) = \frac{1}{b}\,\sin\varphi.
    \label{eq:u0-null}
\end{equation}
Inserting Eq.~\eqref{eq:u0-null} into \eqref{eq:VIM-iter-general} yields the
first-order approximation $u_1(\varphi)$, which already suffices to compute
the leading-order bending angle. The deflection angle $\hat\alpha$ is obtained
from the condition
\begin{equation}
    u(\pi + \hat\alpha) = 0,
\end{equation}
using the small-angle expansion $\sin(\pi+\hat\alpha)\simeq -\hat\alpha$ and
retaining terms up to the desired order in the small parameters (mass scale,
charge, deformation parameter $\delta$, etc.).

In summary, Eqs.~\eqref{eq:master-orbit}, \eqref{eq:S-def},
\eqref{eq:VIM-iter-general}, together with the initial profiles
\eqref{eq:u0-timelike} and \eqref{eq:u0-null}, provide a systematic and
unified framework for applying the variational iteration method to the study
of perihelion precession and light deflection in a wide class of static,
spherically symmetric spacetimes characterized by the generic metric
\eqref{eq:metric-generic-r}.

\section{Impulse approximation in a generic static, spherically symmetric spacetime}

The deflection of light by a massive object is a foundational prediction of General Relativity, famously confirmed by Eddington's 1919 solar eclipse expedition. The standard, rigorous method for calculating the deflection angle involves solving the null geodesic equations in the curved spacetime geometry described by the appropriate metric (e.g., the Schwarzschild metric). This process requires the machinery of tensor calculus, including the calculation of Christoffel symbols and the solution of a second-order differential equation \cite{Bartelmann:1999yn,Yarimoto:2024uew}. While this approach is fundamental and exact, it can be mathematically intensive and may obscure the underlying physical mechanism for those new to the subject.

An alternative and highly intuitive approach is the impulse-method derivation. This method treats the gravitational interaction in the weak-field limit as a cumulative impulse that imparts a transverse momentum to the photon as it travels past the massive object. The core idea is to calculate the total change in the component of the photon's momentum perpendicular to its initial direction of travel.

Assuming the deflection is small, we can approximate the photon's trajectory as a straight line (the unperturbed path). We then integrate the transverse component of the gravitational force (or, more accurately, the geodesic deviation) acting on the photon along this path. The final deflection angle $\alpha$ is then simply the ratio of the total transverse momentum acquired, $\Delta p_\perp$, to the initial momentum of the photon, $p$:
\begin{equation}
    \alpha \approx \frac{\Delta p_\perp}{p} = \frac{1}{p} \int_{-\infty}^{\infty} F_\perp(t) \, dt
\end{equation}
This perspective reframes the problem from one of geometry (following a curved path in spacetime) to one of dynamics (a change in momentum due to a force), a concept familiar from classical and quantum mechanics.

The impulse-method derivation is not merely a calculational shortcut; its value lies in the physical insight and pedagogical clarity it offers. The motivation for its use, particularly in introductory contexts and for building physical intuition, is compelling 
\cite{Weinberg:1972kfs,Misner:1973prb,Schutz:1985jx,Pireaux:2004id,2024arXiv240504529K}
. The primary advantage of the impulse method is its clear physical picture. It directly connects the abstract concept of spacetime curvature to the tangible effect of a momentum change. It allows one to visualize the gravitational field pulling the photon sideways as it passes, with the total effect being the sum of these tiny pulls. This framing is often more accessible than the abstract notion of a particle following a straight line (geodesic) in a curved manifold \cite{Misner:1973prb}. For weak gravitational fields, where the deflection angle is small, the impulse method provides the correct leading-order result with significantly less mathematical overhead than the full geodesic calculation. It bypasses the need for Christoffel symbols and solving a nonlinear differential equation, relying instead on a straightforward integration of the Newtonian gravitational force, corrected by a factor of two to account for the effects of spatial curvature. The method serves as an excellent educational bridge between Newtonian physics and General Relativity. It starts from the Newtonian concept of force and impulse and shows how it must be modified to arrive at the correct relativistic result. This step-by-step reasoning can help students appreciate the distinct contributions of time dilation (the Shapiro delay component) and space curvature to the total deflection. The impulse approximation frames gravitational lensing as a scattering problem, analogous to the scattering of charged particles in an electromagnetic field. This perspective connects General Relativity to the broader language of scattering theory used extensively in quantum mechanics and particle physics, highlighting the universal nature of such concepts. While the full geodesic formalism is indispensable for precision and strong-field calculations, the impulse method provides a powerful, physically transparent, and mathematically simple derivation for the weak-field deflection of light. Its ability to build intuition makes it an invaluable tool for both students and researchers seeking to understand the physical content of General Relativity.

We now extend impulse calculation to a generic static and spherically symmetric spacetime of the form
\begin{equation}
    \mathrm{d}s^2
    = -\alpha(r,\delta)\,\mathrm{d}t^2
      + \gamma(r,\delta)\,\mathrm{d}r^2
      + \beta(r,\delta)\,\bigl(\mathrm{d}\theta^2 + \sin^2\theta\,\mathrm{d}\varphi^2\bigr),
    \label{eq:gen_metric}
\end{equation}
where the parameter $\delta$ controls deviations from the standard GR case.  We assume asymptotic flatness,
\begin{equation}
    \alpha(r,\delta)\xrightarrow[r\to\infty]{} 1,
    \qquad
    \gamma(r,\delta)\xrightarrow[r\to\infty]{} 1,
    \qquad
    \beta(r,\delta)\xrightarrow[r\to\infty]{} r^2,
\end{equation}
so that the spacetime approaches Minkowski at large radius.

In the weak--field regime relevant for light bending at large impact parameter,
we write the lapse function as
\begin{equation}
    \alpha(r,\delta)
    = 1 + \sum_{n\geq 1}\frac{a_n(\delta)}{r^n},
    \label{eq:alpha_expand}
\end{equation}
with coefficients $a_n(\delta)$ that encode the mass monopole, charge--like terms, and any higher multipolar/deformation contribution driven by~$\delta$.
At the order we are working, the spatial sector can always be cast, in suitable \emph{quasi-Cartesian} coordinates $(x,y,z)$, into a conformally flat form
\begin{equation}
    g_{tt} = -\bigl(1 - 2\Phi(r,\delta)\bigr),
    \qquad
    g_{ij} = \bigl(1 + 2\Phi(r,\delta)\bigr)\,\delta_{ij},
    \label{eq:weak_field_cart}
\end{equation}
where $r=\sqrt{x^2+y^2+z^2}$, and the effective potential is related to $\alpha$ by
\begin{equation}
    \Phi(r,\delta)
    = -\frac{1}{2}\bigl[\alpha(r,\delta)-1\bigr]
    = -\frac{1}{2}\sum_{n\geq 1}\frac{a_n(\delta)}{r^n}.
    \label{eq:Phi_from_alpha}
\end{equation}
For the Reissner-Nordstr\"om-like example considered above, one has
\(\alpha(r) = f(r) = 1 - 2m/r + (Q^2+Q_s^2)/r^2\), which indeed gives
\(\Phi(r) = m/r - (Q^2+Q_s^2)/(2r^2)\).

The passage from the generic $(r,\theta,\varphi)$ form~\eqref{eq:gen_metric} to the quasi-Cartesian weak--field form~\eqref{eq:weak_field_cart} is a standard $1$PN gauge choice: to the leading order relevant for our impulse calculation, all the information on the light bending is captured by the single scalar potential~$\Phi(r,\delta)$ extracted from~$\alpha$ via~\eqref{eq:Phi_from_alpha}.  Corrections due to the detailed structure of $\gamma(r,\delta)$ and $\beta(r,\delta)$ enter only at higher post-Newtonian order and are neglected here.

We choose Cartesian coordinates so that the photon travels initially along the $x$--axis with impact parameter $b$ in the $y$--direction and remains in the equatorial plane $z=0$:
\begin{equation}
    x = \lambda,\qquad
    y = b,\qquad
    z = 0,
    \qquad
    p^x = \frac{\mathrm{d}x}{\mathrm{d}\lambda} = \text{const},\quad
    p^y \ll p^x,
\end{equation}
where $\lambda$ is an affine parameter.  For the weak--field metric~\eqref{eq:weak_field_cart}, the Newtonian-type transverse kick may be represented by the spatial
connection term
\begin{equation}
    \Gamma^y{}_{xx}
    = -\partial_y \Phi(r,\delta)
    = -\Phi'(r,\delta)\,\frac{y}{r}
    = -\Phi'(r,\delta)\,\frac{b}{\sqrt{x^2+b^2}},
    \label{eq:Gamma_gen}
\end{equation}
where $\Phi'(r,\delta) = \partial\Phi/\partial r$ and $r=\sqrt{x^2+b^2}$ along the reference trajectory.

The geodesic equation $p^a\nabla_a p^y=0$ gives
\begin{equation}
    \frac{\mathrm{d}p^y}{\mathrm{d}\lambda}
    = -\Gamma^y{}_{xx}\,(p^x)^2
    = \Phi'(r,\delta)\,\frac{b}{r}\,(p^x)^2.
    \label{eq:geod_y_gen}
\end{equation}
Using $x\simeq p^x \lambda$ (straight-line approximation), we may trade $\lambda$ for $x$ via $\mathrm{d}\lambda = \mathrm{d}x/p^x$ and integrate from $x=-\infty$ to $x=+\infty$,
\begin{equation}
    \Delta p^y
    = \int_{-\infty}^{+\infty}\frac{\mathrm{d}p^y}{\mathrm{d}\lambda}\,\mathrm{d}\lambda
    = p^x\,b\int_{-\infty}^{+\infty}
      \frac{\Phi'(r,\delta)}{r}\,\mathrm{d}x,
    \qquad r=\sqrt{x^2+b^2}.
    \label{eq:Dpy_gen}
\end{equation}
For a null ray, $p^x\simeq E$ is approximately constant and equal to the photon energy, so that the small-angle deflection is
\begin{equation}
    \alpha_{\rm imp}(b,\delta)
    \simeq \frac{|\Delta p^y|}{p^x}
    = b\left|\int_{-\infty}^{+\infty}
      \frac{\Phi'(r,\delta)}{r}\,\mathrm{d}x\right|
    = 2b\left|\int_{0}^{+\infty}
      \frac{\Phi'(r,\delta)}{r}\,\mathrm{d}x\right|.
    \label{eq:alpha_imp_gen_integral}
\end{equation}
Equation~\eqref{eq:alpha_imp_gen_integral} is the generic extension of the impulse formula: the bending angle is determined by a one-dimensional line-of-sight integral of the effective potential~$\Phi(r,\delta)$ derived from the lapse~$\alpha(r,\delta)$.

Using the expansion~\eqref{eq:Phi_from_alpha},
\begin{equation}
    \Phi(r,\delta)
    = \sum_{n\geq 1}\frac{\Phi_n(\delta)}{r^n},
    \qquad
    \Phi_n(\delta)
    = -\frac{1}{2}a_n(\delta),
    \label{eq:Phi_series}
\end{equation}
we obtain
\begin{equation}
    \Phi'(r,\delta)
    = -\sum_{n\geq 1}n\,\Phi_n(\delta)\,\frac{1}{r^{n+1}}.
\end{equation}
Inserting this into~\eqref{eq:alpha_imp_gen_integral}, and using $r^2=x^2+b^2$, the contribution of each $n$--th term is
\begin{equation}
    \alpha_{\rm imp}^{(n)}(b,\delta)
    = 2n\,\Phi_n(\delta)\,b\int_0^{+\infty}
      \frac{\mathrm{d}x}{(x^2+b^2)^{(n+2)/2}}.
\end{equation}
The standard integral
\begin{equation}
    \int_0^{+\infty}
    \frac{\mathrm{d}x}{(x^2+b^2)^k}
    = \frac{\sqrt{\pi}}{2}\,
      \frac{\Gamma\!\left(k-\tfrac{1}{2}\right)}
           {\Gamma(k)}\,
      \frac{1}{b^{2k-1}},
    \qquad k> \frac{1}{2},
\end{equation}
with $k=(n+2)/2$, yields
\begin{equation}
    \alpha_{\rm imp}^{(n)}(b,\delta)
    = \frac{n\sqrt{\pi}\,\Phi_n(\delta)}{b^n}\,
      \frac{\Gamma\!\left(\tfrac{n+1}{2}\right)}
           {\Gamma\!\left(\tfrac{n+2}{2}\right)}.
\end{equation}
Summing over all $n\geq 1$, the generic impulse deflection angle takes the compact form
\begin{equation}
    \alpha_{\rm imp}(b,\delta)
    = \sum_{n\geq 1}
      \frac{A_n(\delta)}{b^n},
    \qquad
    A_n(\delta)
    = \frac{n\sqrt{\pi}\,\Phi_n(\delta)}{\,}\,
      \frac{\Gamma\!\left(\tfrac{n+1}{2}\right)}
           {\Gamma\!\left(\tfrac{n+2}{2}\right)}.
    \label{eq:alpha_imp_series}
\end{equation}
In terms of the original coefficients $a_n(\delta)$ in~\eqref{eq:alpha_expand}, this becomes
\begin{equation}
    A_n(\delta)
    = -\frac{n\sqrt{\pi}}{2}\,a_n(\delta)\,
      \frac{\Gamma\!\left(\tfrac{n+1}{2}\right)}
           {\Gamma\!\left(\tfrac{n+2}{2}\right)}.
    \label{eq:A_n_from_a_n}
\end{equation}
For the first two terms one finds explicitly
\begin{align}
    A_1(\delta) &= -a_1(\delta), \\
    A_2(\delta) &= -\frac{\pi}{2}\,a_2(\delta),
\end{align}
so that
\begin{equation}
    \alpha_{\rm imp}(b,\delta)
    = -\frac{a_1(\delta)}{b}
      -\frac{\pi}{2}\,\frac{a_2(\delta)}{b^2}
      + \mathcal{O}\!\left(\frac{1}{b^3}\right).
    \label{eq:alpha_imp_leading}
\end{equation}

Equation~\eqref{eq:alpha_imp_leading} is the raw impulse estimate. For the
Schwarzschild monopole \(a_1=-2m\), it gives \(2m/b\), i.e. one half of
the full relativistic coefficient. When the impulse method is used as a
GR-calibrated estimate, only the monopole coefficient is doubled:
\begin{equation}
    \alpha_{\rm imp,cal}(b,\delta)
    =
    -\frac{2a_1(\delta)}{b}
    -\frac{\pi}{2}\frac{a_2(\delta)}{b^2}
    +\mathcal{O}(b^{-3}).
\end{equation}
This calibrated expression is useful for comparison, but the coefficient
of the \(a_2/b^2\) term should not be regarded as the exact relativistic
coefficient.

\section{Scalar hairy black holes in Einstein-Maxwell-conformally coupled scalar theory}

The classical no-hair theorem posits that black holes in general relativity (GR) are uniquely characterized by just three quantities: mass, electric charge, and angular momentum, and cannot support additional "hair" such as scalar fields outside the event horizon~\cite{Ruffini:1971bza}. This restriction arises from the fact that, unlike electromagnetic and gravitational fields, minimally coupled scalar fields do not obey a Gauss law, rendering scalar hair apparently forbidden in standard GR~\cite{Herdeiro:2015waa}. However, this understanding has been challenged by solutions found in modified theories of gravity, particularly those involving non-minimal or conformal couplings between gravity and scalar fields. A pivotal development in this direction was the discovery of the Bocharova-Bronnikov-Melnikov-Bekenstein (BBMB) black hole, which arises when a scalar field is conformally coupled to gravity. The BBMB solution famously provides a counterexample to the no-hair conjecture, although the scalar field diverges at the horizon~\cite{Bocharova:1970skc,Bekenstein:1974sf}. Further studies established that, within Einstein gravity with a conformally coupled scalar, the BBMB black hole is the unique static solution~\cite{Xanthopoulos:1991mx}. More recently, extensions involving higher curvature terms or additional couplings have been shown to give rise to black holes with primary or secondary scalar hair~\cite{Myung:2019adj}.

Interest in scalarized black holes has surged with the realization that spontaneous scalarization can occur in various extended gravity models, such as those including Gauss-Bonnet or Maxwell terms coupled non-minimally to scalar fields~\cite{Doneva:2017bvd,Silva:2017uqg,Antoniou:2017acq,Herdeiro:2018wub}. In these scenarios, linear perturbation analysis often predicts the onset of instability for the standard black holes, which leads to the emergence of new branches of scalarized (and often charged) black hole solutions, labeled by the number of nodes in the scalar field profile~\cite{Myung:2018vug}. A particularly fruitful setting for exploring such phenomena is the Einstein-Maxwell-conformally coupled scalar (EMCS) theory, which combines quadratic scalar coupling to the electromagnetic sector with conformal coupling to gravity~\cite{Astorino:2013sfa,Astorino:2013xc}. In the case where the scalar-Maxwell coupling vanishes, the EMCS theory admits both the constant scalar hairy black hole and the charged BBMB black hole as solutions, with the former being stable under generic perturbations and the latter remaining unstable due to its extremal character~\cite{Bronnikov:1978mx}. For nonzero coupling, the theory supports a rich spectrum of scalarized black holes, whose properties and stability depend sensitively on the coupling parameter~\cite{Myung:2018vug}.

Black holes with scalar hair, once considered to be excluded in GR, have emerged as natural solutions in a wide class of modified gravity models. These developments have significant implications for our understanding of gravity in strong-field regimes and motivate further study into the structure, stability, and phenomenology of scalarized black holes~\cite{Bekenstein:1974sf,Doneva:2017bvd,Silva:2017uqg,Antoniou:2017acq,Herdeiro:2018wub,Astorino:2013sfa,Astorino:2013xc}.

We begin by deriving the Einstein equations following the formalism presented in \cite{Zou:2019ays}:
\begin{equation} \label{nequa1}
G_{\mu\nu}=2(1+\bar{\alpha} \phi^2) T^{\rm M}_{\mu\nu}+T^{\rm \phi}_{\mu\nu},
\end{equation}
with the Einstein tensor defined conventionally as $G_{\mu\nu}=R_{\mu\nu}-Rg_{\mu\nu}/2$. The corresponding energy-momentum tensors for Maxwell's electrodynamics and a conformally coupled scalar field are expressed as:
\begin{eqnarray} \label{equa2}
T^{\rm M}_{\mu\nu}=F_{\mu\rho}F_{\nu}~^\rho- \frac{F^2}{4}g_{\mu\nu},\label{trace} \end{eqnarray} 
\begin{eqnarray}
T^{\rm \phi}_{\mu\nu}=\hat{\beta}\Big[\phi^2G_{\mu\nu}+g_{\mu\nu}\nabla^2(\phi^2)-\nabla_\mu\nabla_\nu(\phi^2)+6\nabla_\mu\phi\nabla_\nu\phi-3(\nabla\phi)^2g_{\mu\nu}\Big],
\end{eqnarray}
where the Maxwell tensor satisfies the traceless condition $T^{{\rm M} \mu}_\mu=0$. Additionally, the modified Maxwell equations are:
\begin{equation} \label{maxwell-eq}
\nabla^\mu F_{\mu\nu}=2\bar{\alpha} \phi \nabla^{\mu}\phi F^2.
\end{equation}

%\begin{equation}\nabla_\mu\left[\left(1+\bar{\alpha}\phi^2\right)F^{\mu\nu}\right]= ,\end{equation}

The scalar field obeys the wave equation:
\begin{equation} \label{ascalar-eq}
\nabla^2\phi-\frac{1}{6}R\phi-\frac{\bar{\alpha}}{6\hat{\beta
}}  F^2 \phi=0.
\end{equation}
By computing the trace of the Einstein equation \eqref{nequa1} and employing \eqref{ascalar-eq}, the Ricci scalar emerges as:
\begin{equation} \label{ricciz}
R=-\bar{\alpha} \phi^2  F^2.
\end{equation}
When $\bar{\alpha}=0$, we recover $R=0$, consistent with the scenario in EMCS theory without the additional coupling \cite{Bekenstein:1974sf,Xanthopoulos:1991mx}. This occurs as the scalar equation reduces to the standard conformally coupled form $\nabla^2\phi-R\phi/6=0$, despite  $T^{\phi}_{\mu\nu}$ not being traceless. Utilizing Eq.\eqref{ricciz}, the scalar field equation transforms into:
\begin{equation} \label{scalar-eq}
\nabla^2\phi+\frac{\bar{\alpha}}{6}\Big[\phi^2-\frac{1}{\hat{\beta}}\Big]F^2 \phi=0.
\end{equation}

In the limiting case $\bar{\alpha}=0$, Eq.\eqref{scalar-eq} reduces straightforwardly to the minimally coupled scalar field equation   $\nabla^2\phi=0$, which aligns with previously established results \cite{Bekenstein:1974sf,Xanthopoulos:1991mx}. In this particular scenario ($\bar{\alpha}=0$ EMCS theory), the equations of motion simplify to:
\begin{equation} \label{al0-eqs}
G_{\mu\nu}=2T^{\rm M}_{\mu\nu}+T^{\rm \phi}_{\mu\nu},~~\nabla^\mu F_{\mu\nu}=0,~~R=0,~~\nabla^2\phi=0.
\end{equation} These equations admit a static, spherically symmetric hairy charged black hole solution \cite{Astorino:2013sfa,Zou:2019ays}:
\begin{eqnarray}
&&ds^2_{\rm csh}=\bar{g}_{\mu\nu}dx^\mu dx^\nu=-N(r)dt^2+\frac{dr^2}{N(r)}+r^2d\Omega^2_2, \nonumber \\
&&N(r)=1-\frac{2m}{r}+\frac{Q^2+Q^2_s}{r^2},~~\bar{\phi}_c=\pm \sqrt{\frac{1}{\hat{\beta}}}\sqrt{\frac{Q^2_s}{Q^2_s+Q^2}},~~\bar{A}_t=\frac{Q}{r}-\frac{Q}{r_+}. \label{bbmb2}
\end{eqnarray}
This solution follows from the background equations $\bar{G}_{\mu\nu}=2\bar{T}^{\rm M}_{\mu\nu}/(1-\hat{\beta} \bar{\phi}_c^2)=2\bar{T}_{\mu\nu}$, with the background stress-energy tensor $\bar{T}^{\mu}_\nu=\frac{Q^2+Q^2_s}{r^4}{\rm diag}[-1,-1,1,1]$ and a constant scalar field $\bar{\phi}_c={\rm const}$. Hereafter, an overbar indicates background solutions. The parameter $\hat{\beta}$ is related to gravitational constants as  $\hat{\beta}=\kappa/6=4\pi G/3$. The horizon radii are located at  $r_{\pm}=m\pm \sqrt{m^2-Q^2-Q^2_s}$. Notably, the scalar field constitutes a primary scalar hair, slightly altering the horizon positions compared to a standard Reissner-Nordstr\"om (RN) black hole. Nevertheless, since the scalar solution $\bar{\phi}_c$ is coupled with both charges, it does not strictly represent a purely primary hair. We examine a generalization of the Reissner-Nordstr\"om geometry incorporating an additional scalar charge, with the metric represented as:
\begin{equation}
ds^2 = -f(r)dt^2 + \frac{dr^2}{f(r)} + r^2(d\theta^2 + \sin^2\theta\, d\varphi^2),
\end{equation}
where
\begin{equation}
f(r) = 1 - \frac{2m}{r} + \frac{Q^2 + Q_s^2}{r^2}.
\end{equation}
and the parameters $m$, $Q$, and $Q_s$ denote the mass, electric charge, and scalar charge, respectively.

\subsection{Homotopy perturbation method}

 Given its robustness and versatility, HPM is an ideal candidate for calculating the deflection angle in the modified spacetimes of scalar hairy. Its ability to handle complex nonlinearities without restrictive assumptions makes it a powerful tool for exploring new physics and obtaining precise analytical results that can be compared with future astronomical observations.

In the $\bar{\alpha}=0$ limit of the Einstein--Maxwell--conformally coupled
scalar (EMCS) theory, the constant-scalar-hair charged black hole solution
\eqref{bbmb2} can be written in the standard static, spherically symmetric form
\begin{equation}
ds^2 = -f(r)\,dt^2 + \frac{dr^2}{f(r)} 
      + r^2\left(d\theta^2 + \sin^2\theta\,d\varphi^2\right),
\label{eq:metric_csh_HPM}
\end{equation}
with
\begin{equation}
f(r) = 1 - \frac{2m}{r} + \frac{Q^2 + Q_s^2}{r^2}.
\label{eq:f_scalar_hairy}
\end{equation}
It is convenient to introduce an \emph{effective electro-scalar charge}
\begin{equation}
q^2 \equiv Q^2 + Q_s^2,
\end{equation}
so that $f(r) = 1 - 2m/r + q^2/r^2$ has exactly the same functional form as
the Reissner--Nordstr\"om (RN) metric, while the physical interpretation of
$q^2$ now includes the scalar hair.

We consider null geodesics in the equatorial plane, $\theta = \pi/2$. The
conserved energy and angular momentum per unit affine parameter are
\begin{equation}
E = f(r)\,\frac{dt}{d\lambda}, 
\qquad
L = r^2\frac{d\varphi}{d\lambda},
\end{equation}
and the null condition $ds^2=0$ yields
\begin{equation}
\left(\frac{dr}{d\lambda}\right)^2 = 
E^2 - f(r)\,\frac{L^2}{r^2}.
\end{equation}
Introducing the impact parameter $l \equiv L/E$ and the inverse radius
$u(\varphi) \equiv 1/r(\varphi)$, one finds
\begin{equation}
\left(\frac{du}{d\varphi}\right)^2 
= \frac{1}{l^2} - f(u)\,u^2.
\end{equation}
Differentiating with respect to $\varphi$ and using
$f(u) = 1 - 2m u + q^2 u^2$, one obtains the second-order null geodesic
equation
\begin{equation}
\frac{d^2u}{d\varphi^2} 
= -\frac{1}{2}u^2\frac{df}{du} - f(u)\,u
= -u + 3m u^2 - 2q^2 u^3.
\end{equation}
Equivalently,
\begin{equation}
\frac{d^2u}{d\varphi^2} + u = 3m u^2 - 2q^2 u^3,
\label{eq:geodesic_scalar_hairy_HPM}
\end{equation}
which coincides with the RN null geodesic equation~\cite{Shchigolev:2016gro}
upon the replacement $Q^2 \to q^2 = Q^2 + Q_s^2$.

%\subsubsection{Homotopy construction and perturbative solution}

Following the homotopy perturbation method (HPM)
of~\cite{Shchigolev:2016gro}, we rewrite
Eq.~\eqref{eq:geodesic_scalar_hairy_HPM} in the form
\begin{equation}
u'' + u - p\left(3m u^2 - 2q^2 u^3\right) = 0,
\qquad
p \in [0,1],
\label{eq:homotopy_scalar_hairy}
\end{equation}
where the prime denotes derivative with respect to $\varphi$. For $p=0$ we
recover the straight-line equation $u''+u=0$, while for $p=1$ we recover the
full nonlinear equation~\eqref{eq:geodesic_scalar_hairy_HPM}.

We seek a solution in the form of a series in the embedding parameter $p$,
\begin{equation}
u(\varphi) = u_0(\varphi) + p\,u_1(\varphi) + p^2 u_2(\varphi) + \cdots,
\label{eq:u_series_scalar_hairy}
\end{equation}
and impose asymptotically straight-line initial conditions
\begin{equation}
u_0(0)=0,\quad u_0'(0)=\frac{1}{l}, 
\qquad
u_n(0)=u_n'(0)=0,\quad n\geq 1.
\label{eq:ICs_scalar_hairy}
\end{equation}
Substituting~\eqref{eq:u_series_scalar_hairy} into
Eq.~\eqref{eq:homotopy_scalar_hairy} and matching coefficients of $p^n$ gives
the hierarchy
\begin{align}
\mathcal{O}(p^0):\quad & u_0'' + u_0 = 0, 
\label{eq:u0_scalar_hairy}\\
\mathcal{O}(p^1):\quad & u_1'' + u_1 = 3m u_0^2 - 2q^2 u_0^3,
\label{eq:u1_scalar_hairy}\\
\mathcal{O}(p^2):\quad & u_2'' + u_2 = 6m u_0 u_1 - 6q^2 u_0^2 u_1,
\quad\ldots
\end{align}
and so on. For the weak deflection regime it is sufficient to retain only
$u_0$ and $u_1$.

The zeroth-order equation~\eqref{eq:u0_scalar_hairy} with
initial conditions~\eqref{eq:ICs_scalar_hairy} yields the straight-line
trajectory
\begin{equation}
u_0(\varphi) 
= \frac{1}{b}\sin\varphi.
\label{eq:u0_solution_scalar_hairy}
\end{equation}

At first order, Eq.~\eqref{eq:u1_scalar_hairy} becomes
\begin{equation}
u_1'' + u_1
= 3m\frac{\sin^2\varphi}{b^2} 
  - 2q^2\frac{\sin^3\varphi}{b^3},
\end{equation}
with $u_1(0)=u_1'(0)=0$. Using standard trigonometric identities and
solving the linear inhomogeneous ODE for the RN case), one obtains
\begin{equation}
u_1(\varphi)
= \frac{m}{b^2}\big(1-\cos\varphi\big)^2
  - \frac{q^2}{4b^3}\Big[
     \big(\cos^2\varphi+2\big)\sin\varphi - 3\varphi\cos\varphi
   \Big].
\label{eq:u1_solution_scalar_hairy}
\end{equation}
Therefore, the first-order HPM approximation to the null geodesic is
\begin{equation}
u(\varphi) \approx \frac{1}{b}\sin\varphi
+ \frac{m}{b^2}\big(1-\cos\varphi\big)^2
- \frac{q^2}{4b^3}\Big[
    \big(\cos^2\varphi+2\big)\sin\varphi - 3\varphi\cos\varphi
  \Big],
\label{eq:u_approx_scalar_hairy}
\end{equation}
with $q^2 = Q^2 + Q_s^2$. Setting $Q_s=0$ reproduces the RN result, while $q^2\to 0$ recovers the Schwarzschild
case.

The total deflection angle $\hat{\alpha}$ (here denoted by $\beta$) is defined
as the deviation of the outgoing asymptote from the flat-space value
$\varphi=\pi$. Writing
\begin{equation}
\varphi_\infty = \pi + \beta,
\qquad
|\beta|\ll 1,
\end{equation}
the condition that the photon returns to infinity is
\begin{equation}
u(\varphi=\pi+\beta) = 0.
\end{equation}
Using the small-angle approximations
\begin{equation}
\sin(\pi+\beta) \simeq -\beta,
\qquad
\cos(\pi+\beta) \simeq -1,
\end{equation}
and the first-order trajectory~\eqref{eq:u_approx_scalar_hairy}, we find
\begin{align}
u(\pi+\beta) 
&\simeq \frac{1}{b}\sin(\pi+\beta)
 + \frac{m}{b^2}\big(1-\cos(\pi+\beta)\big)^2
 - \frac{q^2}{4b^3}
   \Big[
\big(\cos^2(\pi+\beta)+2\big)\sin(\pi+\beta) 
     - 3(\pi+\beta)\cos(\pi+\beta)
   \Big]
\nonumber\\[1ex]
&\simeq -\frac{\beta}{b}
       + \frac{4m}{b^2}
       - \frac{3\pi q^2}{4b^3},
\end{align}
where terms of order $\mathcal{O}(\beta^2)$ and higher have been neglected.
Imposing $u(\pi+\beta)=0$ yields the weak deflection angle
\begin{equation}
\beta \;\simeq\; \frac{4m}{b} 
                 - \frac{3\pi}{4}\,\frac{q^2}{b^2}
                 \;=\; \frac{4m}{b} 
                 - \frac{3\pi}{4}\,\frac{Q^2+Q_s^2}{b^2}.
\label{eq:deflection_scalar_hairy}
\end{equation}
Thus, the constant scalar hair enters the bending angle exclusively through
the combination $Q^2+Q_s^2$, and \emph{reduces} the deflection compared to
the Schwarzschild value, acting effectively as an additional repulsive charge.
For $Q_s\to 0$, Eq.~\eqref{eq:deflection_scalar_hairy} reduces to the
standard RN expression~\cite{Shchigolev:2016gro},
\begin{equation}
\beta_{\rm RN} \simeq \frac{4m}{b} 
 - \frac{3\pi}{4}\,\frac{Q^2}{b^2}.
\end{equation}

Higher-order corrections in $m/l$ and $q^2/l^2$ can be obtained systematically
either by including the next HPM term $u_2$ in
Eq.~\eqref{eq:u_series_scalar_hairy}, or by employing the HPM-based root
expansion for the deflection equation $u(\varphi)=0$ as in
Ref.~\cite{Shchigolev:2016gro}.

\subsection{Variational iteration method}
%Its effectiveness has been demonstrated in computing light deflection and perihelion precession in various spherically symmetric spacetimes, showing excellent agreement with established results. This success provides strong motivation for extending VIM to more complex or modified gravity scenarios-such as Kerr, Reissner-Nordstr\"om, or scalar-tensor black holes-where analytic solutions are often elusive.

By incorporating VIM into our study, we gain a semi-analytical method founded on variational principles, which serves as a valuable cross-check for results obtained by other techniques. This not only enhances the robustness of our findings for the deflection angle in scalar hairy and bumblebee black hole spacetimes, but also enables a broader comparative assessment of leading approximation methods in gravitational lensing and modified gravity.

We consider a static, spherically symmetric spacetime with line element
\begin{equation}
  \mathrm{d}s^2
  = -f(r)\,\mathrm{d}t^2 + f(r)^{-1}\,\mathrm{d}r^2 + r^2\,\left(\mathrm{d}\theta^2+\sin^2\theta\,\mathrm{d}\varphi^2\right),
\end{equation}
where the lapse function is
\begin{equation}
  f(r) \;=\; 1 - \frac{2m}{r} + \frac{Q^2+Q_s^2}{r^2}\,.
\end{equation}
For null geodesics in the equatorial plane, introducing $u(\varphi)=1/r(\varphi)$ yields
\begin{equation}
  \frac{\mathrm{d}^2u}{\mathrm{d}\varphi^2} + u
  = 3m\,u^2 \;-\; 2\,(Q^2+Q_s^2)\,u^3.
  \label{eq:orbit2}
\end{equation}

Define the linear operator 
$L[u] = u'' + u,
  \qquad
  N[u] = -\,3m\,u^2 + 2\,(Q^2+Q_s^2)\,u^3,
$
so that \eqref{eq:orbit2} reads $L[u]+N[u]=0$.  The VIM correction functional is \cite{Shchigolev:2016vpz}
\begin{equation}
  u_{n+1}(\varphi)
  = u_n(\varphi)
  + \int_{0}^{\varphi}
    \lambda(\phi)\,\left[L[u_n(\phi)] + N[\tilde u_n(\phi)]\right]\,
    \mathrm{d}\phi,
\end{equation}
where $\tilde u_n$ is held fixed under variation.  Imposing
$\delta u_{n+1}=0$ fixes the Lagrange multiplier via
$
\lambda''(\phi)+\lambda(\phi)=0,
  \quad
  \lambda(\varphi)=0,
  \quad
  \lambda'(\varphi)=1
  \;\;\Longrightarrow\;\;
  \lambda(\phi)=\sin(\phi-\varphi).$ We choose the straight-line approximation
$u_0(\varphi)=\frac{\sin\varphi}{b},$
with $b$ the impact parameter.  Then the first iterate is
\begin{equation}
  u_1(\varphi)
  = \frac{\sin\varphi}{b}
  - \frac{3m}{b^2}
    \int_{0}^{\varphi}\!\sin(\phi-\varphi)\,\sin^2\phi\,\mathrm{d}\phi
  + \frac{2(Q^2+Q_s^2)}{b^3}
    \int_{0}^{\varphi}\!\sin(\phi-\varphi)\,\sin^3\phi\,\mathrm{d}\phi.
\end{equation}
Evaluating these integrals gives
\begin{align}
  u_1(\varphi)
  &= \frac{\sin\varphi}{b}
   + \frac{m}{b^2}\,(1-\cos\varphi)^2
   + \frac{Q^2+Q_s^2}{4b^3}
      \left(3\varphi\cos\varphi+\sin^3\varphi-3\sin\varphi\right).
\end{align}
The total deflection angle $\alpha$ is determined by $u(\pi+\alpha)=0$. Expanding for small
$m/b$ and $(Q^2+Q_s^2)/b^2$ yields
\begin{equation}
  \alpha = \frac{4m}{b}
          \;-\;\frac{3\pi\,(Q^2+Q_s^2)}{4\,b^2}
          + \mathcal{O}\left(b^{-3}\right),
\end{equation}
which reproduces the Schwarzschild result for $Q=Q_s=0$ and gives the leading
charge correction characteristic of Reissner-Nordstr\"om-type spacetimes. The obtained result is consistent with the findings reported in Ref.~\cite{QiQi:2023nex}.

\subsection{Impulse method derivation}

We study the trajectory of a light ray in the spacetime described by
\begin{equation}\label{eq:metric}
    \mathrm{d}s^2
    = -f(r)\,\mathrm{d}t^2 + f(r)^{-1}\,\mathrm{d}r^2 + r^2\left(\mathrm{d}\theta^2+\sin^2\theta\,\mathrm{d}\varphi^2\right)\,,
\end{equation}
with the lapse function
\begin{equation}\label{eq:lapse}
    f(r)
    = 1 - \frac{2m}{r} + \frac{Q^2 + Q_s^2}{r^2}\,,
\end{equation}
where \(m\) is the central mass, \(Q\) the electric charge, and \(Q_s\) a secondary scalar charge.

In Cartesian coordinates \((x,y,z)\) with \(r=\sqrt{x^2+y^2+z^2}\), we write the metric to leading order in \(m/r\) and \((Q^2+Q_s^2)/r^2\) as
\begin{align}
    g_{tt} &= -\left(1 - 2\Phi(r)\right)\,, \nonumber \\ 
    g_{ij} &= \left(1 + 2\Phi(r)\right)\,\delta_{ij}\,, \nonumber \\
    \Phi(r) &= \frac{m}{r} - \frac{Q^2 + Q_s^2}{2\,r^2}\,.
\end{align}
Here \(\Phi(r)\) plays the role of the Newtonian potential.

We choose coordinates so that the photon travels initially along the \(x\)-axis, at fixed impact parameter \(b\) in the \(y\)-direction, and stays in the equatorial plane \(z=0\).  Thus
$
    x = \lambda,\quad y = b,\quad z = 0,
    \qquad
    p^x = \frac{\mathrm{d}x}{\mathrm{d}\lambda} = \mathrm{const}, 
    \quad
    p^y \ll p^x,
$
where \(\lambda\) is an affine parameter. The only nonvanishing Christoffel symbol contributing to the bending in the \(y\)-direction is
\begin{equation}\label{eq:Gamma}
    \Gamma^y{}_{xx}
    = -\partial_y\Phi(r)
    = -\Phi'(r)\,\frac{y}{r}
    = \frac{m\,y}{r^3} \;-\;\frac{(Q^2+Q_s^2)\,y}{r^4}\,.
\end{equation}
The geodesic equation \(p^a\nabla_a p^y=0\) then gives
\begin{equation}\label{eq:geod_y}
    \frac{\mathrm{d}p^y}{\mathrm{d}\lambda}
    = -\Gamma^y{}_{xx}\,(p^x)^2
    = -\frac{m\,b}{(x^2+b^2)^{3/2}}\,(p^x)^2
      + \frac{(Q^2+Q_s^2)\,b}{(x^2+b^2)^2}\,(p^x)^2.
\end{equation}

Integrating from $x=-\infty$ to $x=+\infty$, and using
\[
\int_{-\infty}^{+\infty}\frac{b\,dx}{(x^2+b^2)^{3/2}}=\frac{2}{b},
\qquad
\int_{-\infty}^{+\infty}\frac{b\,dx}{(x^2+b^2)^2}=\frac{\pi}{2b^2},
\]
we obtain
\begin{equation}
\Delta p^y
=
(p^x)^2\left[
-\frac{2m}{b}
+\frac{\pi\,(Q^2+Q_s^2)}{2b^2}
\right].
\end{equation}
Hence the \emph{raw} impulse estimate for the deflection angle is
\begin{equation}
\alpha_{\rm imp,raw}
=
\frac{|\Delta p^y|}{p^x}
\simeq
\frac{2m}{b}
-\frac{\pi\,(Q^2+Q_s^2)}{2b^2}
+\mathcal{O}(b^{-3}).
\label{eq:alpha_imp_raw_scalar}
\end{equation}

As is well known, the naive impulse treatment reproduces only one half of the
full Schwarzschild coefficient, since it captures the Newtonian-like transverse
kick but not the full relativistic contribution from spatial curvature.
To compare on the same footing with the geodesic-based weak-field result, one
may therefore introduce a simple GR calibration of the monopole term, leading to
\begin{equation}
\alpha_{\rm imp,cal}
\simeq
\frac{4m}{b}
-\frac{\pi\,(Q^2+Q_s^2)}{2b^2}
+\mathcal{O}(b^{-3}).
\label{eq:alpha_imp_cal_scalar}
\end{equation}
The first expression, Eq.~\eqref{eq:alpha_imp_raw_scalar}, is the direct outcome
of the impulse derivation, while Eq.~\eqref{eq:alpha_imp_cal_scalar} is a
phenomenologically calibrated version used only for comparison with the
standard relativistic weak-deflection coefficient.

Above equation shows that the leading bending angle \(4m/b\) is reduced by a term proportional to the square of the total charge. In full general relativity the exact Reissner--Nordstr\"om result exhibits a different coefficient on the charge-squared term: HPM/VIM (and the standard PPN treatment) yield the factor $-3\pi(Q^2+Q_s^2)/(4b^2)$, whereas the impulse approximation gives $-\pi(Q^2+Q_s^2)/(2b^2)$. This discrepancy is expected: the impulse method uses a Newtonian-type potential plus a single GR-inspired correction factor, so it is reliable for the leading Schwarzschild term $4m/b$ and for the \emph{sign} and scaling of the charge correction, but not for its precise relativistic coefficient.

{\color{black}\subsection{Numerical results near the photon sphere}
\label{sec:numerical-benchmark}
 
To delineate the practical range of validity of the three semi-analytical
schemes derived above, we now compare them against the exact null-geodesic
bending integral for the scalar-hairy Reissner--Nordstr\"om-like geometry
\eqref{eq:f_scalar_hairy}. The photon-sphere radius and the associated
critical impact parameter are obtained from the standard conditions
$2f(r_{\rm ph}) = r_{\rm ph}\,f'(r_{\rm ph})$ and $b_{\rm ph} = r_{\rm ph}/\sqrt{f(r_{\rm ph})}$,
giving
\begin{equation}
r_{\rm ph} \;=\; \frac{3m + \sqrt{9m^{2} - 8q^{2}}}{2},
\qquad
b_{\rm ph} \;=\; \frac{r_{\rm ph}}{\sqrt{f(r_{\rm ph})}},
\qquad q^{2}\equiv Q^{2}+Q_{s}^{2}.
\label{eq:rph_bph_scalarhairy}
\end{equation}
 
The exact bending angle is computed from the master integral
\eqref{eq:exact_bending_generic} specialized to $\alpha=f$, $\gamma=1/f$,
$\beta=r^{2}$,
\begin{equation}
\hat\alpha_{\rm ex}(b)
\;=\;
2\!\int_{r_{0}(b)}^{\infty}
\frac{b\,dr}
     {r\,\sqrt{r^{2}-b^{2}f(r)}}
\;-\;\pi,
\label{eq:alpha_exact_scalarhairy}
\end{equation}
where the closest-approach radius $r_{0}(b)$ is the largest root of
$r_{0}^{2}=b^{2}f(r_{0})$. The integrable square-root singularity at
$r=r_{0}$ is regularized by the substitution $r=r_{0}+u^{2}$ before
numerical quadrature. For comparison we use the leading weak-field
predictions
\begin{align}
\hat\alpha_{\rm HPM/VIM}(b) &\;=\; \frac{4m}{b}-\frac{3\pi\,q^{2}}{4\,b^{2}},
\label{eq:alpha_hpmvim_scalarhairy}
\\[2pt]
\hat\alpha_{\rm imp}(b) &\;=\; \frac{4m}{b}-\frac{\pi\,q^{2}}{2\,b^{2}},
\label{eq:alpha_imp_scalarhairy}
\end{align}
where Eq.~\eqref{eq:alpha_imp_scalarhairy} corresponds to the calibrated
impulse approximation~\eqref{eq:alpha_imp_cal_scalar}. At the order
retained here the HPM and VIM bending series coincide and therefore appear
as a single curve in all figures.
 
As a representative non-Schwarzschild example we choose $q=0.8\,m$, for
which Eq.~\eqref{eq:rph_bph_scalarhairy} gives $r_{\rm ph}\simeq 2.4849\,m$
and $b_{\rm ph}\simeq 4.5460\,m$. Figure~\ref{fig:benchmark_combined}(a)
shows $\hat\alpha$ versus $b/b_{\rm ph}$ on a logarithmic scale over the
broad interval $b_{\rm ph}\le b\le 8\,b_{\rm ph}$, while
Figure~\ref{fig:benchmark_combined}(b) displays the corresponding relative
errors. Three quantitative trends emerge. First, all three weak-field
approximations underestimate the exact bending once $b$ approaches
$b_{\rm ph}$ from above, reflecting the well-known logarithmic enhancement
of the deflection in the strong-lensing regime. Second, the relative error
decreases monotonically as $b/b_{\rm ph}$ grows, falling from about $\sim
80\%$ at $b\simeq 1.05\,b_{\rm ph}$ to a few percent at
$b\gtrsim 5\,b_{\rm ph}$. Third, the impulse and HPM/VIM curves
bracket the exact result and differ from each other by less than one
percentage point, with the impulse value happening to lie marginally
closer in this particular interval. This is essentially coincidental: at
$q=0.8\,m$ the Schwarzschild monopole $4m/b$ dominates over the charge
correction, and the larger (more negative) HPM/VIM coefficient
$-3\pi/4$ pulls its prediction slightly further below the exact result
than does the impulse coefficient $-\pi/2$. Outside this window the
HPM/VIM expression is preferable for precision weak-lensing work because
it reproduces the correct relativistic coefficient of the $q^{2}/b^{2}$
term, whereas the impulse coefficient is smaller by exactly a factor of
$2/3$ and therefore underestimates the charge correction by $33\%$ at
fixed $q^{2}/b^{2}$.

\begin{table}[t]
\centering
\caption{{\color{black}Complementary numerical comparison at fixed impact parameter $b=20\,m$
for several values of the effective charge $q=\sqrt{Q^{2}+Q_{s}^{2}}$. The
column $b/b_{\rm ph}$ uses the charge-dependent critical impact parameter
$b_{\rm ph}(q)$ from Eq.~\eqref{eq:rph_bph_scalarhairy}. Symbols are:
$\hat\alpha_{\rm ex}$ is the exact deflection from numerical quadrature of
Eq.~\eqref{eq:alpha_exact_scalarhairy}, $\hat\alpha_{\rm HPM/VIM}$ is the
common leading HPM/VIM prediction \eqref{eq:alpha_hpmvim_scalarhairy}, and
$\hat\alpha_{\rm imp}$ is the calibrated impulse estimate
\eqref{eq:alpha_imp_scalarhairy}. The last two columns give the relative errors
with respect to the exact result.}}
\begin{tabular}{ccccccc}
\toprule
$q/m$ & $b/b_{\rm ph}$ & $\hat\alpha_{\rm ex}$ &
$\hat\alpha_{\rm HPM/VIM}$ & $\hat\alpha_{\rm imp}$ &
$\Delta_{\rm HPM/VIM}(\%)$ & $\Delta_{\rm imp}(\%)$ \\
\midrule
0.0 & 3.8490 & 0.2361 & 0.2000 & 0.2000 & 15.303 & 15.303 \\
0.2 & 3.8750 & 0.2358 & 0.1998 & 0.1998 & 15.278 & 15.245 \\
0.4 & 3.9581 & 0.2347 & 0.1991 & 0.1994 & 15.204 & 15.070 \\
0.6 & 4.1163 & 0.2330 & 0.1979 & 0.1986 & 15.079 & 14.776 \\
0.8 & 4.3995 & 0.2306 & 0.1962 & 0.1975 & 14.905 & 14.360 \\
1.0 & 5.0000 & 0.2275 & 0.1941 & 0.1961 & 14.682 & 13.819 \\
\bottomrule
\end{tabular}
\label{tab:benchmark_charge_dependence}
\end{table}

Third, the relative error of the leading HPM/VIM expression with respect
to the exact result is at the \(15\%\) level for the fixed choice
\(b=20m\). This error is dominated by pure Schwarzschild post-Minkowskian
terms omitted from the leading weak-field expression. Including the
standard second-order Schwarzschild contribution \(15\pi m^{2}/(4b^{2})\)
reduces the residual error to the few-percent level. Therefore the table \ref{tab:benchmark_charge_dependence}
should be interpreted as a test of the leading HPM/VIM charge coefficient,
not as a complete high-accuracy weak-field expansion.

\begin{figure*}[t]
    \centering
    \includegraphics[width=0.95\textwidth]{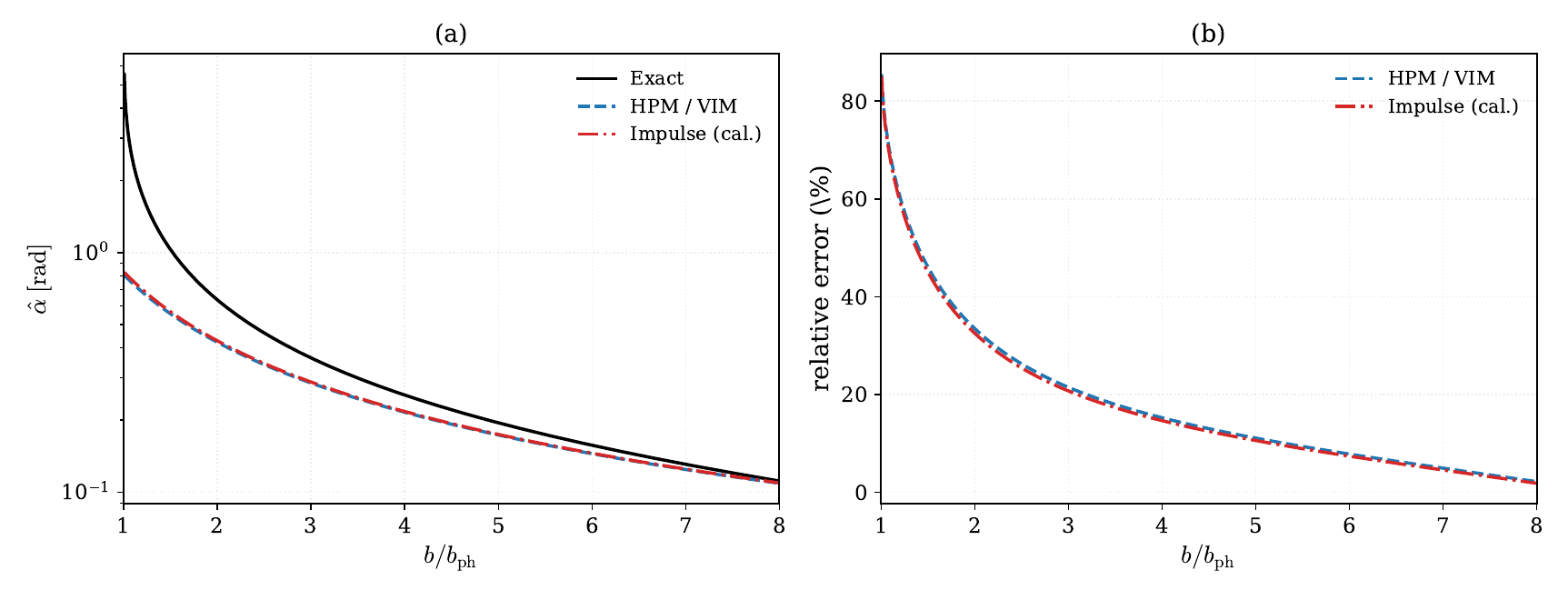}
    \caption{{\color{black}For the scalar-hairy Reissner--Nordstr\"om-like
geometry with $q=0.8\,m$. Panel (a): bending angle $\hat\alpha$ as a
function of $b/b_{\rm ph}$ on a logarithmic scale. The solid curve is the
exact result obtained from numerical quadrature of
Eq.~\eqref{eq:alpha_exact_scalarhairy}; the dashed and dot-dashed curves
correspond, respectively, to the leading HPM/VIM prediction
\eqref{eq:alpha_hpmvim_scalarhairy} and to the calibrated impulse
estimate \eqref{eq:alpha_imp_scalarhairy}. As $b\to b_{\rm ph}^{+}$ the
exact deflection diverges logarithmically while the weak-field
approximations remain finite. Panel (b): relative error of the
semi-analytical predictions with respect to the exact result. Both
approximations exhibit a smooth, monotonic decay of the relative error
from $\mathcal{O}(70\%)$ near the photon sphere to a few percent at
$b/b_{\rm ph}\gtrsim 5$.}}
    \label{fig:benchmark_combined}
\end{figure*}
 
To exhibit the role of the effective charge $q^{2}=Q^{2}+Q_{s}^{2}$
across the parameter space, Fig.~\ref{fig:family_q} shows the exact
deflection angle as a function of $b/m$ for several values of $q/m$
ranging from the Schwarzschild case ($q/m=0$) to the extremal limit
($q/m=1$). Two physical effects are visible. (i) At fixed $b/m$,
increasing $q/m$ \emph{reduces} the bending, in agreement with the sign
of the $q^{2}$ correction in
Eqs.~\eqref{eq:alpha_hpmvim_scalarhairy}--\eqref{eq:alpha_imp_scalarhairy}:
the scalar/electric charge contributes a repulsive component to the
effective potential. (ii) The leftmost endpoint of each curve, set by the
critical impact parameter $b_{\rm ph}(q)$ defined in
Eq.~\eqref{eq:rph_bph_scalarhairy}, moves inward as $q/m$ grows. The
photon sphere itself shrinks from $r_{\rm ph}=3\,m$ at $q=0$ down to
$r_{\rm ph}=2\,m$ at $q=m$, while $b_{\rm ph}$ decreases from $3\sqrt{3}\,m
\simeq 5.196\,m$ to $4\,m$. Thus, for a given measured bending angle, an
allowed range of $q$ corresponds to a non-trivial joint constraint on
both $r_{\rm ph}$ and $b_{\rm ph}$, which is precisely the type of
``lensing fingerprint'' that the weak-field expansion makes accessible in
closed form.
 
\begin{figure}[t]
    \centering
    \includegraphics[width=0.78\textwidth]{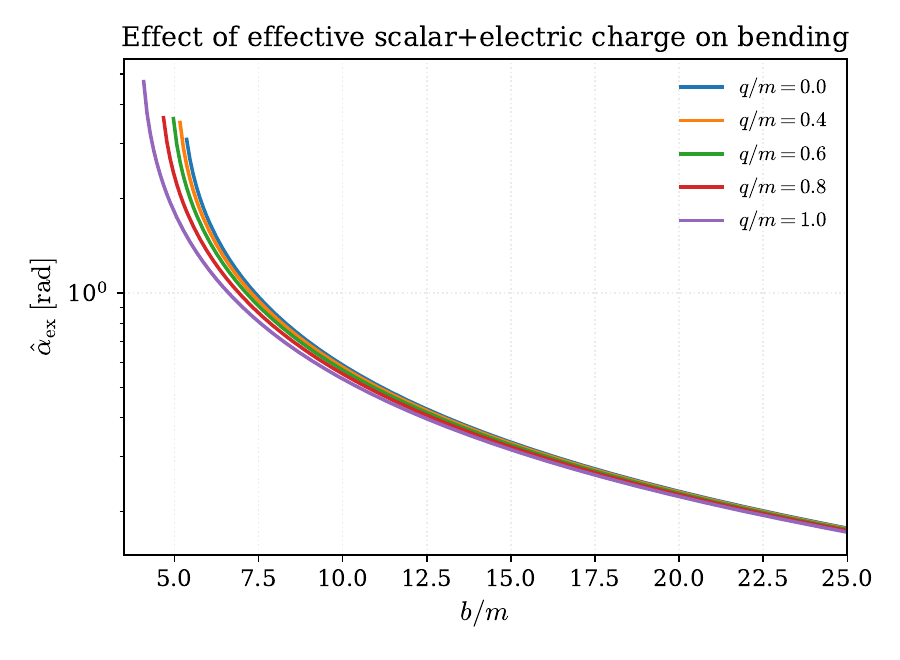}
    \caption{{\color{black}Effect of the effective charge $q=\sqrt{Q^{2}+Q_{s}^{2}}$ on
the exact bending angle. Each curve corresponds to a different value of
$q/m$, with the leftmost endpoint set by the critical impact parameter
$b_{\rm ph}(q)$ of Eq.~\eqref{eq:rph_bph_scalarhairy}. At fixed $b/m$,
increasing $q/m$ reduces the deflection, consistent with the sign of the
$q^{2}/b^{2}$ correction in
Eqs.~\eqref{eq:alpha_hpmvim_scalarhairy}--\eqref{eq:alpha_imp_scalarhairy}.
The Schwarzschild curve ($q/m=0$) is the upper envelope; the extremal
case $q/m=1$ corresponds to the lower curve, with $r_{\rm ph}=2\,m$ and
$b_{\rm ph}=4\,m$.}}
    \label{fig:family_q}
\end{figure}

The combined picture from Figs.~\ref{fig:benchmark_combined}--\ref{fig:family_q}
 can be summarized as follows.
For $b\gtrsim 3\,b_{\rm ph}$ the leading HPM/VIM expression is in
excellent agreement with the exact bending angle and is the
recommended choice for precision weak-lensing applications, since it
reproduces both the Schwarzschild monopole $4m/b$ and the
relativistically correct charge coefficient $-3\pi q^{2}/(4b^{2})$. The
calibrated impulse approximation is essentially indistinguishable from
the HPM/VIM prediction in this regime and provides a transparent
physical interpretation of the bending in terms of an effective potential
$\Phi(r,\delta)$, but underestimates the charge coefficient by $33\%$
and is therefore inferior for high-precision parameter inference. For
$1.5\,b_{\rm ph}\lesssim b\lesssim 3\,b_{\rm ph}$ all three weak-field
approximations remain qualitatively informative but acquire
$\mathcal{O}(20\%\!-\!50\%)$ errors. }

\section{Conclusion}
In this work we have developed a unified semi-analytical framework for weak
gravitational lensing in a generic static, spherically symmetric spacetime of
the form
\(
ds^2 = -\alpha(r,\delta)\,dt^2 + \gamma(r,\delta)\,dr^2 +
\beta(r,\delta)\,d\Omega_2^2,
\)
where the functions $\alpha$, $\gamma$ and $\beta$ encode both the standard
Schwarzschild sector and deviations associated with scalar hair. Starting from the exact
first-order null geodesic equation, we derived a compact master orbit
equation for $u(\varphi)=1/r(\varphi)$ and a corresponding generic
expression for the deflection angle $\hat\alpha(b)$ in terms of the metric
functions and their derivatives. This provides a model-independent ``front end'': once a specific solution
is supplied via \(\alpha(r,\delta)\), \(\gamma(r,\delta)\), and
\(\beta(r,\delta)\), the corresponding weak-lensing observables can be
computed systematically.

We then solved the master orbit equation by three independent
semi-analytical methods: the homotopy perturbation method (HPM), the
variational iteration method (VIM), and a calibrated impulse (single-kick)
approximation formulated in terms of an effective potential. HPM and VIM
both yield rapidly convergent series solutions without requiring an explicit
small physical parameter in the original differential equation; the relevant
expansion parameters emerge naturally as $m/b$, charge to impact parameter
ratios. For the examples considered, we found exact agreement between HPM
and VIM for the weak deflection series, including the leading order
general relativistic terms and the first non-trivial corrections associated
with scalar hair. This mutual consistency, together
with agreement with known PPN/generalized-PPN expansions from the literature,
demonstrates that HPM and VIM provide a robust and flexible semi-analytical
toolbox for lensing in generic static spacetimes.

As concrete test beds, we applied the generic formalism to the physically
motivated geometry of a scalar-hairy black hole in Einstein-Maxwell-conformally
coupled scalar theory, where the metric function depends on an effective
charge. In the scalar-hairy case we obtained the standard
Schwarzschild term $4m/b$ together with charge-dependent corrections, thereby isolating the imprint of scalar
hair on the weak deflection angle. These ``lensing fingerprints'' quantify, in closed
form, how scalar hair deforms the standard GR
predictions in the weak-field regime. We then complemented the analytic derivation with a direct numerical results near the photon sphere, showing explicitly how the exact bending departs from all first-order weak-field approximations aas \(b\to b_{\rm ph}^{+}\).

The impulse (transverse-kick) method, recast within the same generic
framework by expanding the lapse into an effective potential, reproduces
the leading Schwarzschild deflection and correctly captures the sign and
scaling of the non-Schwarzschild corrections. As expected from its
essentially Newtonian character, some higher-order coefficients differ
from the fully relativistic HPM/VIM results, but the impulse method
remains an efficient and physically transparent tool for obtaining quick
estimates and building intuition, especially when exploring large
parameter spaces or when only leading-order trends are required.

Our analysis highlights several directions for future work. On the
theoretical side, the generic semi-analytical machinery developed here
can be extended to strong-deflection and photon-sphere regimes, to
time-delay and magnification observables, and to non-spherically symmetric
cases such as slowly rotating or axisymmetric deformations of Kerr-like
metrics. On the phenomenological side, the closed-form deflection
coefficients obtained for scalar-hairy and bumblebee black holes can be
combined with realistic lens models and current or forthcoming
astrophysical data (e.g. EHT, VLBI, or high-precision microlensing) to
place quantitative constraints on scalar charge
parameter. More broadly, the generic $(\alpha,\gamma,\beta)$ formulation,
together with HPM, VIM, and the impulse approximation, offers a versatile
and computationally economical framework for mapping a wide variety of
modified gravity and hairy black hole solutions into observable lensing
signatures and for using those signatures to test gravity beyond General
Relativity.

\acknowledgments
 A. \"O and R. P. would like to acknowledge networking support of the COST Action CA21106 - COSMIC WISPers in the Dark Universe: Theory, astrophysics and experiments (CosmicWISPers), the COST Action CA22113 - Fundamental challenges in theoretical physics (THEORY-CHALLENGES), the COST Action CA21136 - Addressing observational tensions in cosmology with systematics and fundamental physics (CosmoVerse), the COST Action CA23130 - Bridging high and low energies in search of quantum gravity (BridgeQG), and the COST Action CA23115 - Relativistic Quantum Information (RQI) funded by COST (European Cooperation in Science and Technology).  A. \"O and R. P. would also like to acknowledge the funding support of SCOAP3.

 \section*{Data Availability Statement}
Data sharing not applicable to this article as no datasets were generated or analysed during the current study.

\section*{Declaration}
During the preparation of this work the author used [ChatGPT by OpenAI] in order to improve language of the paper. After using this tool/service, the author reviewed and edited the content as needed and takes full responsibility for the content of the published article.

\bibliography{ref}

@article{Shchigolev:2015sgg,
    author = "Shchigolev, V. K.",
    title = "{Analytical Computation of the Perihelion Precession in General Relativity via the Homotopy Perturbation Method}",
    eprint = "1706.01809",
    archivePrefix = "arXiv",
    primaryClass = "gr-qc",
    doi = "10.13189/ujcmj.2015.030401",
    journal = "J. Comput. Math.",
    volume = "3",
    pages = "45--49",
    year = "2015"
}

@article{Shchigolev:2016vpz,
    author = "Shchigolev, V. K.",
    title = "{Variational iteration method for studying perihelion precession and deflection of light in General Relativity}",
    doi = "10.14419/ijpr.v4i2.6530",
    journal = "Int. J. Phys. Res.",
    volume = "4",
    number = "2",
    pages = "52",
    year = "2016"
}

@article{2024arXiv240504529K,
	title = {Gravitational {Deflection} of {Light}: {A} {Heuristic} {Derivation} at the {Undergraduate} {Level}},
	volume = {74},
	issn = {0374-4914},
	url = {http://www.npsm-kps.org/journal/view.html?doi=10.3938/NPSM.74.394},
	doi = {10.3938/NPSM.74.394},
	number = {4},
archivePrefix = {arXiv},
       eprint = {2405.04529},
	journal = {New Physics: Sae Mulli},
	author = {Kim, Hongbin and Yeom, Dong-han and Kim, Jong Hyun},
	month = apr,
	year = {2024},
	note = {Edition: 2024/04/30
Publisher: New Physics: Sae Mulli},
	pages = {394--400},
}

@article{Gibbons:2008hb,
    author = "Gibbons, G. W. and Warnick, C. M.",
    title = "{Universal properties of the near-horizon optical geometry}",
    eprint = "0809.1571",
    archivePrefix = "arXiv",
    primaryClass = "gr-qc",
    reportNumber = "DAMTP-2008-80",
    doi = "10.1103/PhysRevD.79.064031",
    journal = "Phys. Rev. D",
    volume = "79",
    pages = "064031",
    year = "2009"
}

@article{HE20073,
title = {Variational iteration method-Some recent results and new interpretations},
journal = {Journal of Computational and Applied Mathematics},
volume = {207},
number = {1},
pages = {3-17},
year = {2007},
note = {Special Issue: Variational Iteration Method-Reality, Potential, and Challenges},
issn = {0377-0427},
doi = {https://doi.org/10.1016/j.cam.2006.07.009},
url = {https://www.sciencedirect.com/science/article/pii/S0377042706004559},
author = {Ji-Huan He},
}

@misc{Huang:2025vqm,
    author = "Huang, Yang and Fu, Xiangyun and Lu, Zhenyan and Qin, Xin",
    title = "{The influence of cosmological constant on light deflection in rotating spacetimes via the generalized Gibbons-Werner method}",
    eprint = "2505.06999",
    archivePrefix = "arXiv",
    primaryClass = "gr-qc",
    month = "5",
    year = "2025"
}

@article{Huang:2023bto,
    author = "Huang, Yang and Cao, Zhoujian and Lu, Zhenyan",
    title = "{Generalized Gibbons-Werner method for stationary spacetimes}",
    eprint = "2306.04145",
    archivePrefix = "arXiv",
    primaryClass = "gr-qc",
    doi = "10.1088/1475-7516/2024/01/013",
    journal = "JCAP",
    volume = "01",
    pages = "013",
    year = "2024"
}

@article{Li:2019mqw,
    author = "Li, Zonghai and Zhou, Tao",
    title = "{Equivalence of Gibbons-Werner method to geodesics method in the study of gravitational lensing}",
    eprint = "1908.05592",
    archivePrefix = "arXiv",
    primaryClass = "gr-qc",
    doi = "10.1103/PhysRevD.101.044043",
    journal = "Phys. Rev. D",
    volume = "101",
    number = "4",
    pages = "044043",
    year = "2020"
}

@article{WAZWAZ2007895,
title = {The variational iteration method for solving linear and nonlinear systems of PDEs},
journal = {Computers and Mathematics with Applications},
volume = {54},
number = {7},
pages = {895-902},
year = {2007},
note = {Variational Iteration Method for Nonlinear Problems},
issn = {0898-1221},
doi = {https://doi.org/10.1016/j.camwa.2006.12.059},
url = {https://www.sciencedirect.com/science/article/pii/S0898122107003033},
author = {Abdul-Majid Wazwaz},
}

@book{Misner:1973prb,
    author = "Misner, Charles W. and Thorne, K. S. and Wheeler, J. A.",
    title = "{Gravitation}",
    isbn = "978-0-7167-0344-0, 978-0-691-17779-3",
    publisher = "W. H. Freeman",
    address = "San Francisco",
    year = "1973"
}

@book{Schutz:1985jx,
    author = "Schutz, Bernard F.",
    title = "{A FIRST COURSE IN GENERAL RELATIVITY}",
    doi = "10.1017/CBO9780511984181",
    isbn = "978-0-511-98418-1",
    publisher = "Cambridge Univ. Pr.",
    address = "Cambridge, UK",
    year = "1985"
}

@article{Pireaux:2004id,
    author = "Pireaux, Sophie",
    title = "{Light deflection in Weyl gravity: Critical distances for photon paths}",
    eprint = "gr-qc/0403071",
    archivePrefix = "arXiv",
    doi = "10.1088/0264-9381/21/7/011",
    journal = "Class. Quant. Grav.",
    volume = "21",
    pages = "1897--1913",
    year = "2004"
}

@book{Weinberg:1972kfs,
    author = "Weinberg, Steven",
    title = "{Gravitation and Cosmology}: {Principles and Applications of the General Theory of Relativity}",
    isbn = "978-0-471-92567-5, 978-0-471-92567-5",
    publisher = "John Wiley and Sons",
    address = "New York",
    year = "1972"
}

@article{QiQi:2023nex,
    author = "Qi, Qi and Meng, Yuan and Wang, Xi-Jing and Kuang, Xiao-Mei",
    title = "{Gravitational lensing effects of black hole with conformally coupled scalar hair}",
    doi = "10.1140/epjc/s10052-023-12233-z",
    journal = "Eur. Phys. J. C",
    volume = "83",
    number = "11",
    pages = "1043",
    year = "2023"
}

@article{Bartelmann:1999yn,
    author = "Bartelmann, Matthias and Schneider, Peter",
    title = "{Weak gravitational lensing}",
    eprint = "astro-ph/9912508",
    archivePrefix = "arXiv",
    doi = "10.1016/S0370-1573(00)00082-X",
    journal = "Phys. Rept.",
    volume = "340",
    pages = "291--472",
    year = "2001"
}

@article{Yarimoto:2024uew,
    author = "Yarimoto, Hirotaka and Oguri, Masamune",
    title = "{Born approximation in wave optics of gravitational lensing revisited}",
    eprint = "2412.07272",
    archivePrefix = "arXiv",
    primaryClass = "astro-ph.CO",
    doi = "10.1103/PhysRevD.111.083541",
    journal = "Phys. Rev. D",
    volume = "111",
    number = "8",
    pages = "083541",
    year = "2025"
}

@article{Will:2014kxa,
    author = "Will, Clifford M.",
    title = "{The Confrontation between General Relativity and Experiment}",
    eprint = "1403.7377",
    archivePrefix = "arXiv",
    primaryClass = "gr-qc",
    doi = "10.12942/lrr-2014-4",
    journal = "Living Rev. Rel.",
    volume = "17",
    pages = "4",
    year = "2014"
}

@article{Pantig:2025deu,
    author = {Pantig, Reggie C. and {\"O}vg{\"u}n, Ali},
    title = "{Gravitational black hole shadow spectroscopy}",
    eprint = "2509.05594",
    archivePrefix = "arXiv",
    primaryClass = "hep-th",
    doi = "10.1103/jmpd-8tn8",
    journal = "Phys. Rev. D",
    volume = "112",
    number = "12",
    pages = "124072",
    year = "2025"
}

@book{Schneider:1992bmb,
    author = {Schneider, Peter and Ehlers, J{\"u}rgen and Falco, Emilio E.},
    title = "{Gravitational Lenses}",
    doi = "10.1007/978-3-662-03758-4",
    isbn = "978-3-540-66506-9, 978-3-662-03758-4",
    publisher = "Springer",
    series = "Astronomy and Astrophysics Library",
    year = "1992"
}

@inproceedings{Narayan:1996ba,
    author = "Narayan, Ramesh and Bartelmann, Matthias",
    title = "{Lectures on gravitational lensing}",
    booktitle = "{13th Jerusalem Winter School in Theoretical Physics: Formation of Structure in the Universe}",
    eprint = "astro-ph/9606001",
    archivePrefix = "arXiv",
    month = "6",
    year = "1996"
}

@article{Virbhadra:1999nm,
    author = "Virbhadra, K. S. and Ellis, George F. R.",
    title = "{Schwarzschild black hole lensing}",
    eprint = "astro-ph/9904193",
    archivePrefix = "arXiv",
    doi = "10.1103/PhysRevD.62.084003",
    journal = "Phys. Rev. D",
    volume = "62",
    pages = "084003",
    year = "2000"
}

@article{Bozza:2010xqn,
    author = "Bozza, Valerio",
    title = "{Gravitational Lensing by Black Holes}",
    eprint = "0911.2187",
    archivePrefix = "arXiv",
    primaryClass = "gr-qc",
    doi = "10.1007/s10714-010-0988-2",
    journal = "Gen. Rel. Grav.",
    volume = "42",
    pages = "2269--2300",
    year = "2010"
}

@article{Bozza:2001xd,
    author = "Bozza, V. and Capozziello, S. and Iovane, G. and Scarpetta, G.",
    title = "{Strong field limit of black hole gravitational lensing}",
    eprint = "gr-qc/0102068",
    archivePrefix = "arXiv",
    doi = "10.1023/A:1012292927358",
    journal = "Gen. Rel. Grav.",
    volume = "33",
    pages = "1535--1548",
    year = "2001"
}

@article{Virbhadra:2008ws,
    author = "Virbhadra, K. S.",
    title = "{Relativistic images of Schwarzschild black hole lensing}",
    eprint = "0810.2109",
    archivePrefix = "arXiv",
    primaryClass = "gr-qc",
    doi = "10.1103/PhysRevD.79.083004",
    journal = "Phys. Rev. D",
    volume = "79",
    pages = "083004",
    year = "2009"
}

@article{Virbhadra:2002ju,
    author = "Virbhadra, K. S. and Ellis, G. F. R.",
    title = "{Gravitational lensing by naked singularities}",
    doi = "10.1103/PhysRevD.65.103004",
    journal = "Phys. Rev. D",
    volume = "65",
    pages = "103004",
    year = "2002"
}

@article{Claudel:2000yi,
    author = "Claudel, Clarissa-Marie and Virbhadra, K. S. and Ellis, G. F. R.",
    title = "{The Geometry of photon surfaces}",
    eprint = "gr-qc/0005050",
    archivePrefix = "arXiv",
    doi = "10.1063/1.1308507",
    journal = "J. Math. Phys.",
    volume = "42",
    pages = "818--838",
    year = "2001"
}

@article{Virbhadra:1998dy,
    author = "Virbhadra, K. S. and Narasimha, D. and Chitre, S. M.",
    title = "{Role of the scalar field in gravitational lensing}",
    eprint = "astro-ph/9801174",
    archivePrefix = "arXiv",
    journal = "Astron. Astrophys.",
    volume = "337",
    pages = "1--8",
    year = "1998"
}

@article{Virbhadra:2007kw,
    author = "Virbhadra, K. S. and Keeton, C. R.",
    title = "{Time delay and magnification centroid due to gravitational lensing by black holes and naked singularities}",
    eprint = "0710.2333",
    archivePrefix = "arXiv",
    primaryClass = "gr-qc",
    doi = "10.1103/PhysRevD.77.124014",
    journal = "Phys. Rev. D",
    volume = "77",
    pages = "124014",
    year = "2008"
}

@article{Virbhadra:2022iiy,
    author = "Virbhadra, K. S.",
    title = "{Distortions of images of Schwarzschild lensing}",
    eprint = "2204.01879",
    archivePrefix = "arXiv",
    primaryClass = "gr-qc",
    doi = "10.1103/PhysRevD.106.064038",
    journal = "Phys. Rev. D",
    volume = "106",
    number = "6",
    pages = "064038",
    year = "2022"
}

@article{Virbhadra:2024xpk,
    author = "Virbhadra, K. S.",
    title = "{Conservation of distortion of gravitationally lensed images}",
    eprint = "2402.17190",
    archivePrefix = "arXiv",
    primaryClass = "gr-qc",
    doi = "10.1103/PhysRevD.109.124004",
    journal = "Phys. Rev. D",
    volume = "109",
    number = "12",
    pages = "124004",
    year = "2024"
}

@article{Shchigolev:2016gro,
    author = "Shchigolev, V. K. and Bezbatko, D. N.",
    title = "{Studying Gravitational Deflection of Light by Kiselev Black Hole via Homotopy Perturbation Method}",
    eprint = "1612.07279",
    archivePrefix = "arXiv",
    primaryClass = "gr-qc",
    doi = "10.1007/s10714-019-2521-6",
    journal = "Gen. Rel. Grav.",
    volume = "51",
    number = "2",
    pages = "34",
    year = "2019"
}

@article{Keeton:2005jd,
    author = "Keeton, Charles R. and Petters, A. O.",
    title = "{Formalism for testing theories of gravity using lensing by compact objects. I. Static, spherically symmetric case}",
    eprint = "gr-qc/0511019",
    archivePrefix = "arXiv",
    doi = "10.1103/PhysRevD.72.104006",
    journal = "Phys. Rev. D",
    volume = "72",
    pages = "104006",
    year = "2005"
}

@article{Iyer:2009wa,
    author = "Iyer, S. V. and Hansen, E. C.",
    title = "{Light's Bending Angle in the Equatorial Plane of a Kerr Black Hole}",
    eprint = "0907.5352",
    archivePrefix = "arXiv",
    primaryClass = "gr-qc",
    doi = "10.1103/PhysRevD.80.124023",
    journal = "Phys. Rev. D",
    volume = "80",
    pages = "124023",
    year = "2009"
}

@article{Gibbons:2008rj,
    author = "Gibbons, G. W. and Werner, M. C.",
    title = "{Applications of the Gauss-Bonnet theorem to gravitational lensing}",
    eprint = "0807.0854",
    archivePrefix = "arXiv",
    primaryClass = "gr-qc",
    doi = "10.1088/0264-9381/25/23/235009",
    journal = "Class. Quant. Grav.",
    volume = "25",
    pages = "235009",
    year = "2008"
}

@article{Ishihara:2016vdc,
    author = "Ishihara, Asahi and Suzuki, Yusuke and Ono, Toshiaki and Kitamura, Takao and Asada, Hideki",
    title = "{Gravitational bending angle of light for finite distance and the Gauss-Bonnet theorem}",
    eprint = "1604.08308",
    archivePrefix = "arXiv",
    primaryClass = "gr-qc",
    doi = "10.1103/PhysRevD.94.084015",
    journal = "Phys. Rev. D",
    volume = "94",
    number = "8",
    pages = "084015",
    year = "2016"
}

@article{Werner:2012rc,
    author = "Werner, M. C.",
    title = "{Gravitational lensing in the Kerr-Randers optical geometry}",
    eprint = "1205.3876",
    archivePrefix = "arXiv",
    primaryClass = "gr-qc",
    doi = "10.1007/s10714-012-1458-9",
    journal = "Gen. Rel. Grav.",
    volume = "44",
    pages = "3047--3057",
    year = "2012"
}

@article{Li:2020dln,
    author = {Li, Zonghai and {\"O}vg{\"u}n, Ali},
    title = "{Finite-distance gravitational deflection of massive particles by a Kerr-like black hole in the bumblebee gravity model}",
    eprint = "2001.02074",
    archivePrefix = "arXiv",
    primaryClass = "gr-qc",
    doi = "10.1103/PhysRevD.101.024040",
    journal = "Phys. Rev. D",
    volume = "101",
    number = "2",
    pages = "024040",
    year = "2020"
}

@article{he1999,
title = {Homotopy perturbation technique},
journal = {Computer Methods in Applied Mechanics and Engineering},
volume = {178},
number = {3},
pages = {257-262},
year = {1999},
issn = {0045-7825},
doi = {https://doi.org/10.1016/S0045-7825(99)00018-3},
url = {https://www.sciencedirect.com/science/article/pii/S0045782599000183},
author = {Ji-Huan He},
}

@article{he2000,
title = {A coupling method of a homotopy technique and a perturbation technique for non-linear problems},
journal = {International Journal of Non-Linear Mechanics},
volume = {35},
number = {1},
pages = {37-43},
year = {2000},
issn = {0020-7462},
doi = {https://doi.org/10.1016/S0020-7462(98)00085-7},
url = {https://www.sciencedirect.com/science/article/pii/S0020746298000857},
author = {Ji-Huan He},
}

@article{Astorino:2013sfa,
    author = "Astorino, Marco",
    title = "{C-metric with a conformally coupled scalar field in a magnetic universe}",
    eprint = "1307.4021",
    archivePrefix = "arXiv",
    primaryClass = "gr-qc",
    reportNumber = "CECS-PHY-13-05",
    doi = "10.1103/PhysRevD.88.104027",
    journal = "Phys. Rev. D",
    volume = "88",
    number = "10",
    pages = "104027",
    year = "2013"
}

@article{Ruffini:1971bza,
    author = "Ruffini, Remo and Wheeler, John A.",
    title = "{Introducing the black hole}",
    doi = "10.1063/1.3022513",
    journal = "Phys. Today",
    volume = "24",
    number = "1",
    pages = "30",
    year = "1971"
}

@article{Herdeiro:2015waa,
    author = "Herdeiro, Carlos A. R. and Radu, Eugen",
    editor = "Herdeiro, Carlos A. R. and Cardoso, Vitor and Lemos, Jose P. S. and Mena, Filipe C.",
    title = "{Asymptotically flat black holes with scalar hair: a review}",
    eprint = "1504.08209",
    archivePrefix = "arXiv",
    primaryClass = "gr-qc",
    doi = "10.1142/S0218271815420146",
    journal = "Int. J. Mod. Phys. D",
    volume = "24",
    number = "09",
    pages = "1542014",
    year = "2015"
}

@misc{Bocharova:1970skc,
    author = "Bocharova, N. M. and Bronnikov, K. A. and Melnikov, V. N.",
    title = "On one exact solution of the Einstein system of equations and a massless scalar field (Russian)",
    year = "1970"
}

@article{Xanthopoulos:1991mx,
    author = "Xanthopoulos, B. C. and Zannias, T.",
    title = "{The Gravity of three forms}",
    doi = "10.1063/1.529174",
    journal = "J. Math. Phys.",
    volume = "32",
    pages = "2459--2467",
    year = "1991"
}

@article{Myung:2019adj,
    author = "Myung, Yun Soo and Zou, De-Cheng",
    title = "{Black holes in new massive conformal gravity}",
    eprint = "1907.09676",
    archivePrefix = "arXiv",
    primaryClass = "gr-qc",
    doi = "10.1103/PhysRevD.100.064057",
    journal = "Phys. Rev. D",
    volume = "100",
    number = "6",
    pages = "064057",
    year = "2019"
}

@article{Doneva:2017bvd,
    author = "Doneva, Daniela D. and Yazadjiev, Stoytcho S.",
    title = "{New Gauss-Bonnet Black Holes with Curvature-Induced Scalarization in Extended Scalar-Tensor Theories}",
    eprint = "1711.01187",
    archivePrefix = "arXiv",
    primaryClass = "gr-qc",
    doi = "10.1103/PhysRevLett.120.131103",
    journal = "Phys. Rev. Lett.",
    volume = "120",
    number = "13",
    pages = "131103",
    year = "2018"
}

@article{Silva:2017uqg,
    author = "Silva, Hector O. and Sakstein, Jeremy and Gualtieri, Leonardo and Sotiriou, Thomas P. and Berti, Emanuele",
    title = "{Spontaneous scalarization of black holes and compact stars from a Gauss-Bonnet coupling}",
    eprint = "1711.02080",
    archivePrefix = "arXiv",
    primaryClass = "gr-qc",
    doi = "10.1103/PhysRevLett.120.131104",
    journal = "Phys. Rev. Lett.",
    volume = "120",
    number = "13",
    pages = "131104",
    year = "2018"
}

@article{Antoniou:2017acq,
    author = "Antoniou, G. and Bakopoulos, A. and Kanti, P.",
    title = "{Evasion of No-Hair Theorems and Novel Black-Hole Solutions in Gauss-Bonnet Theories}",
    eprint = "1711.03390",
    archivePrefix = "arXiv",
    primaryClass = "hep-th",
    doi = "10.1103/PhysRevLett.120.131102",
    journal = "Phys. Rev. Lett.",
    volume = "120",
    number = "13",
    pages = "131102",
    year = "2018"
}

@article{Herdeiro:2018wub,
    author = "Herdeiro, Carlos A. R. and Radu, Eugen and Sanchis-Gual, Nicolas and Font, Jos{\'e} A.",
    title = "{Spontaneous Scalarization of Charged Black Holes}",
    eprint = "1806.05190",
    archivePrefix = "arXiv",
    primaryClass = "gr-qc",
    doi = "10.1103/PhysRevLett.121.101102",
    journal = "Phys. Rev. Lett.",
    volume = "121",
    number = "10",
    pages = "101102",
    year = "2018"
}

@article{Myung:2018vug,
    author = "Myung, Yun Soo and Zou, De-Cheng",
    title = {{Instability of Reissner{\textendash}Nordstr{\"o}m black hole in Einstein-Maxwell-scalar theory}},
    eprint = "1808.02609",
    archivePrefix = "arXiv",
    primaryClass = "gr-qc",
    doi = "10.1140/epjc/s10052-019-6792-6",
    journal = "Eur. Phys. J. C",
    volume = "79",
    number = "3",
    pages = "273",
    year = "2019"
}

@article{Astorino:2013xc,
    author = "Astorino, Marco",
    title = "{Embedding hairy black holes in a magnetic universe}",
    eprint = "1301.6794",
    archivePrefix = "arXiv",
    primaryClass = "gr-qc",
    reportNumber = "CECS-PHY-13-01",
    doi = "10.1103/PhysRevD.87.084029",
    journal = "Phys. Rev. D",
    volume = "87",
    number = "8",
    pages = "084029",
    year = "2013"
}

@article{Bronnikov:1978mx,
    author = "Bronnikov, K. A. and Kireev, Yu. N.",
    title = "{Instability of Black Holes with Scalar Charge}",
    doi = "10.1016/0375-9601(78)90030-0",
    journal = "Phys. Lett. A",
    volume = "67",
    pages = "95--96",
    year = "1978"
}

@article{Bekenstein:1974sf,
    author = "Bekenstein, J. D.",
    title = "{Exact solutions of Einstein conformal scalar equations}",
    doi = "10.1016/0003-4916(74)90124-9",
    journal = "Annals Phys.",
    volume = "82",
    pages = "535--547",
    year = "1974"
}

@article{Zou:2019ays,
    author = "Zou, De-Cheng and Myung, Yun Soo",
    title = "{Scalar hairy black holes in Einstein-Maxwell-conformally coupled scalar theory}",
    eprint = "1911.08062",
    archivePrefix = "arXiv",
    primaryClass = "gr-qc",
    doi = "10.1016/j.physletb.2020.135332",
    journal = "Phys. Lett. B",
    volume = "803",
    pages = "135332",
    year = "2020"
}

@article{Ditta:2023wye,
    author = "Ditta, Allah and Tiecheng, Xia and Mumtaz, Saadia and Atamurotov, Farruh and Mustafa, G. and Abdujabbarov, Ahmadjon",
    title = "{Testing metric-affine gravity using particle dynamics and photon motion}",
    eprint = "2303.05438",
    archivePrefix = "arXiv",
    primaryClass = "gr-qc",
    doi = "10.1016/j.dark.2023.101248",
    journal = "Phys. Dark Univ.",
    volume = "41",
    pages = "101248",
    year = "2023"
}

@article{Ditta:2023ccf,
    author = "Ditta, Allah and Tiecheng, Xia and Atamurotov, Farruh and Mustafa, G. and Aripov, M. M.",
    title = "{Particle dynamics and weak gravitational lensing around nonlinear electrodynamics black hole}",
    doi = "10.1016/j.cjph.2023.04.018",
    journal = "Chin. J. Phys.",
    volume = "83",
    pages = "664--679",
    year = "2023"
}

@article{Ditta:2023rhr,
    author = "Ditta, Allah and Tiecheng, Xia and Atamurotov, Farruh and Hussain, Ibrar and Mustafa, G.",
    title = "{Particle dynamics, black hole shadow and weak gravitational lensing in the f (Q) theory of gravity}",
    doi = "10.1088/1572-9494/ad0e05",
    journal = "Commun. Theor. Phys.",
    volume = "75",
    number = "12",
    pages = "125404",
    year = "2023"
}

@article{Mustafa:2024zsx,
    author = "Mustafa, G. and Ditta, Allah and Javed, Faisal and Atamurotov, Farruh and Hussain, Ibrar and Ahmedov, Bobomurat",
    title = "{Probing a black hole in Starobinsky-Bel-Robinson gravity with thermodynamical analysis, effective force and gravitational weak lensing}",
    eprint = "2401.08254",
    archivePrefix = "arXiv",
    primaryClass = "gr-qc",
    doi = "10.1016/j.cjph.2024.04.038",
    journal = "Chin. J. Phys.",
    volume = "90",
    pages = "494--508",
    year = "2024"
}

@article{Ashraf:2024dwg,
    author = "Ashraf, Asifa and Ditta, Allah and Sofuo{\u{g}}lu, De{\u{g}}er and Ma, Wen-Xiu and Javed, Faisal and Atamurotov, Farruh and Mahmood, Asif",
    title = "{Quasi-periodic oscillations and particle motion around charged black hole surrounded by a cloud of strings and quintessence field in Rastall gravity}",
    doi = "10.1088/1402-4896/ad3e36",
    journal = "Phys. Scripta",
    volume = "99",
    number = "6",
    pages = "065011",
    year = "2024"
}

@article{Feng:2024jeq,
    author = "Feng, Yihu and Ditta, Allah and Mustafa, G. and Maurya, S. K. and Mahmood, Asif and Atamurotov, Farruh",
    title = "{Non-commutative Schwarzschild black hole surrounded by Perfect fluid dark matter: Plasma lensing and thermodynamics analysis}",
    doi = "10.1016/j.nuclphysb.2024.116713",
    journal = "Nucl. Phys. B",
    volume = "1008",
    pages = "116713",
    year = "2024"
}

@article{Ashraf:2025cnz,
    author = "Ashraf, Asifa and Ditta, Allah and Bouzenada, Abdelmalek and Maurya, S. K. and Abd-Elmonem, Assmaa and Suoliman, Nagat A. A. and Channuie, Phongpichit",
    title = "{Plasma lensing, epicyclic oscillations, particle collision, and thermal fluctuations around a short-hairy black hole}",
    doi = "10.1016/j.dark.2025.101836",
    journal = "Phys. Dark Univ.",
    volume = "48",
    pages = "101836",
    year = "2025"
}

@article{Mushtaq:2025ewk,
    author = "Mushtaq, Farzan and Tiecheng, Xia and Ditta, Allah and Singh, Shalini and Mahmood, Asif",
    title = "{Analyzing the gravitational lensing and epicyclic oscillations around a regular charged black hole}",
    doi = "10.1016/j.nuclphysb.2025.116901",
    journal = "Nucl. Phys. B",
    volume = "1015",
    pages = "116901",
    year = "2025"
}

@article{Ditta:2025bri,
    author = "Ditta, Allah and Mehmood, Sikander and Tanveer, Wajiha and Ahmad, Waleed and Ibraheem, Awad A. and Zulqarnain, Rana Muhammad and Atamurotov, Farruh",
    title = "{Plasma lensing and particle dynamics in the vicinity of a Dyonic black hole solution in the context a perfect fluid}",
    doi = "10.1016/j.nuclphysb.2025.116949",
    journal = "Nucl. Phys. B",
    volume = "1017",
    pages = "116949",
    year = "2025"
}

@article{Rahmatov:2025gpk,
    author = "Rahmatov, Bekzod and Egamberdiev, Islom and Murodov, Sardor and Rayimbaev, Javlon and Ibragimov, Inomjon and Davletov, Erkaboy and Djumanov, Sherzod",
    title = "{Gravitational lensing by black holes surrounded by PFDM in Kalb-Ramond gravity in plasma medium}",
    doi = "10.1016/j.dark.2025.102152",
    journal = "Phys. Dark Univ.",
    volume = "50",
    pages = "102152",
    year = "2025"
}

@article{Rayimbaev:2023bjs,
    author = "Rayimbaev, Javlon and Dialektopoulos, Konstantinos F. and Sarikulov, Furkat and Abdujabbarov, Ahmadjon",
    title = "{Quasiperiodic oscillations around hairy black holes in Horndeski gravity}",
    eprint = "2307.03019",
    archivePrefix = "arXiv",
    primaryClass = "gr-qc",
    doi = "10.1140/epjc/s10052-023-11769-4",
    journal = "Eur. Phys. J. C",
    volume = "83",
    number = "7",
    pages = "572",
    year = "2023"
}

@article{Rayimbaev:2021vsq,
    author = "Rayimbaev, Javlon and Abdujabbarov, Ahmadjon and Wen-Biao, Han",
    title = "{Regular nonminimal magnetic black hole as a source of quasiperiodic oscillations}",
    doi = "10.1103/PhysRevD.103.104070",
    journal = "Phys. Rev. D",
    volume = "103",
    number = "10",
    pages = "104070",
    year = "2021"
}

@article{Rayimbaev:2021luv,
    author = "Rayimbaev, Javlon and Narzilloev, Bakhtiyor and Abdujabbarov, Ahmadjon and Ahmedov, Bobomurat",
    title = "{Dynamics of Magnetized and Magnetically Charged Particles around Regular Nonminimal Magnetic Black Holes}",
    doi = "10.3390/galaxies9040071",
    journal = "Galaxies",
    volume = "9",
    number = "4",
    pages = "71",
    year = "2021"
}

@article{Atamurotov:2015nra,
    author = "Atamurotov, Farruh and Ahmedov, Bobomurat",
    title = "{Optical properties of black hole in the presence of plasma: shadow}",
    eprint = "1507.08131",
    archivePrefix = "arXiv",
    primaryClass = "gr-qc",
    doi = "10.1103/PhysRevD.92.084005",
    journal = "Phys. Rev. D",
    volume = "92",
    pages = "084005",
    year = "2015"
}

@article{Atamurotov:2013sca,
    author = "Atamurotov, Farruh and Abdujabbarov, Ahmadjon and Ahmedov, Bobomurat",
    title = "{Shadow of rotating non-Kerr black hole}",
    doi = "10.1103/PhysRevD.88.064004",
    journal = "Phys. Rev. D",
    volume = "88",
    number = "6",
    pages = "064004",
    year = "2013"
}

@article{Mustafa:2022xod,
    author = {Mustafa, Ghulam and Atamurotov, Farruh and Hussain, Ibrar and Shaymatov, Sanjar and {\"O}vg{\"u}n, Ali},
    title = "{Shadows and gravitational weak lensing by the Schwarzschild black hole in the string cloud background with quintessential field*}",
    eprint = "2207.07608",
    archivePrefix = "arXiv",
    primaryClass = "gr-qc",
    doi = "10.1088/1674-1137/ac917f",
    journal = "Chin. Phys. C",
    volume = "46",
    number = "12",
    pages = "125107",
    year = "2022"
}

@article{Abdujabbarov:2017pfw,
    author = "Abdujabbarov, Ahmadjon and Ahmedov, Bobomurat and Dadhich, Naresh and Atamurotov, Farruh",
    title = "{Optical properties of a braneworld black hole: Gravitational lensing and retrolensing}",
    doi = "10.1103/PhysRevD.96.084017",
    journal = "Phys. Rev. D",
    volume = "96",
    number = "8",
    pages = "084017",
    year = "2017"
}

@article{Atamurotov:2021imh,
    author = "Atamurotov, Farruh and Shaymatov, Sanjar and Sheoran, Pankaj and Siwach, Sanjay",
    title = "{Charged black hole in 4D Einstein-Gauss-Bonnet gravity: particle motion, plasma effect on weak gravitational lensing and centre-of-mass energy}",
    eprint = "2105.02214",
    archivePrefix = "arXiv",
    primaryClass = "gr-qc",
    doi = "10.1088/1475-7516/2021/08/045",
    journal = "JCAP",
    volume = "08",
    pages = "045",
    year = "2021"
}

@article{Atamurotov:2022knb,
    author = {Atamurotov, Farruh and Hussain, Ibrar and Mustafa, Ghulam and {\"O}vg{\"u}n, Ali},
    title = "{Weak deflection angle and shadow cast by the charged-Kiselev black hole with cloud of strings in plasma*}",
    doi = "10.1088/1674-1137/ac9fbb",
    journal = "Chin. Phys. C",
    volume = "47",
    number = "2",
    pages = "025102",
    year = "2023"
}

@article{Atamurotov:2021qds,
    author = "Atamurotov, Farruh and Abdujabbarov, Ahmadjon and Rayimbaev, Javlon",
    title = "{Weak gravitational lensing Schwarzschild-MOG black hole in plasma}",
    doi = "10.1140/epjc/s10052-021-08919-x",
    journal = "Eur. Phys. J. C",
    volume = "81",
    number = "2",
    pages = "118",
    year = "2021"
}

@article{Atamurotov:2022slw,
    author = "Atamurotov, Farruh and Sarikulov, Furkat and Abdujabbarov, Ahmadjon and Ahmedov, Bobomurat",
    title = "{Gravitational weak lensing by black hole in Horndeski gravity in presence of plasma}",
    doi = "10.1140/epjp/s13360-022-02548-3",
    journal = "Eur. Phys. J. Plus",
    volume = "137",
    number = "3",
    pages = "336",
    year = "2022"
}
\end{document}